\documentstyle[preprint,aps]{revtex}

\begin{document}
\draft

\preprint{HD--TVP--95--06}

\title{Hadronization in the \\ $SU(3)$ Nambu--Jona--Lasinio Model}

\author{P.~Rehberg\thanks{E-Mail:
        {\tt Peter@Frodo.TPhys.Uni-Heidelberg.DE}},
        S.~P.~Klevansky and J.~H\"ufner}
\address{Institut f\"ur Theoretische Physik, Universit\"at Heidelberg, \\
         Philosophenweg 19, D--69120 Heidelberg, Germany}

\date{June 1995}

\maketitle
\begin{abstract}
The hadronization process for quarks combining into two mesons, $q\bar
q\to MM'$ at temperature $T$ is described within the $SU(3)$
Nambu--Jona--Lasinio model with finite current quark masses. Invariant
matrix elements, cross--sections and transition rates are calculated to
leading order in a $1/N_c$ expansion.  Four independent classes, $u\bar
d$, $u\bar s$, $u\bar u$ and $s\bar s\to {\rm hadrons}$ are analysed,
and the yield is found to be dominated by pion production.  Threshold
behaviour is determined by the exothermic or endothermic nature of the
processes constituting the hadronization class. A strong suppression of
transition rates is found at the pionic Mott temperature
$T_{M\pi}=212\,{\rm MeV}$, at which the pion becomes a resonant state.
The mean time for hadronization is calculated to be $2\mbox{--}4\,{\rm
fm}/c$ near the Mott temperature.  The calculation of strangeness
changing processes indicates that hadronization accounts for a 1\%
increase in the absolute value of the kaon to pion ratio at
$T=150\,{\rm MeV}$.
\end{abstract}

\pacs{PACS numbers: 11.30Rd, 12.39.Fe, 13.60.Le, 24.85.+p}

\section{Introduction}
One of the current outstanding problems facing nuclear and particle
theoreticians today lies in understanding the phase transition from a
constituent quark and gluonic matter to that of observed hadronic
matter. Since it is expected firstly that such a deconfined state of
quarks and gluons exists, and secondly that it may be observed via
$Pb+Pb$ and other heavy ion collisions already and also to be
undertaken at CERN, an adequate description of the hadronization
process that leads one from the quark and gluon degrees of freedom
inherent to QCD to the observed hadron spectrum is really required. The
present state of the art (for an overview over the field see the
proceedings of the Quark Matter conferences \cite{qm91,qm93} or the
book \cite{hwa}) is characterized either by phenomenological approaches
for the parton to hadron transition \cite{barz,alcor} or computer codes
based on string phenomenology \cite{werner}.

In this paper, we place our emphasis on describing the hadronization
process of quarks and antiquarks into two mesons, within a microscopic,
field theoretical, and non--perturbative framework, which is carried
through at finite temperatures. In itself this is a demanding project
that cannot yet be performed directly starting from the QCD Lagrangian,
and thus the price that we have to pay is in making a choice of a {\em
model\/} Lagrangian. To this end, we invoke the Nambu--Jona--Lasinio
(NJL) model \cite{vogl,spk,hatku} in its $SU_L(3)\times SU_R(3)$
version, which has been constructed to display the same internal
symmetries as QCD itself. This model is known to provide a good
description of the static properties pertaining to both the nonstrange
and strange meson sector at zero temperature, and it allows for a
transparent description at finite temperatures, explicitly displaying
the chiral symmetry restoration phase transition at a critical
temperature of about 200\,MeV in the chiral limit. Although we are
aware of the deficiencies of the model --- lack of confinement as well
as non--renormalizability in a strict field theoretic sense --- it
nevertheless is possible to construct a comprehensive physical picture
{}from the calculations performed to date both at $T=0$ and at finite
temperatures and densities. For definiteness, it is useful to argue
{}from the chiral limit.  Assuming that the chiral and deconfinement
phase transitions occur at the same temperature --- an assumption
supported by lattice gauge simulations\cite{karsch} --- one can build a
heuristic picture on the basis of the NJL model.  At $T\ge T_c$,
hadronic states are unstable, and the system is dominated by
interacting quarks with constituent mass $m=0$.  For $T\le T_c$, where
chiral symmetry is broken, $m\ne 0$, $m_\pi=0$, $m_\sigma=2m$, and a
{\em plasma\/} of quarks and mesons is formed.  While the appearance of
quarks for temperatures $T\le T_c$ is an artifact of the model due to
the absence of confinement, the situation for $T\ge T_c$ is realistic,
except for the lack of explicit gluonic degrees of freedom.  As such,
however, the process of hadronization may be reliable around the
critical temperature, where the shortcomings of the model may not be
too severe. These arguments can be extended to the physical case of
non--zero current quark masses, which we in fact study here.

The purpose of this paper is thus to investigate all hadronization
processes of the type $q\bar q\to MM'$, where $M$ represents a meson in
the non--strange or strange sector, and here $q=u,d,s$. In particular,
we evaluate scattering amplitudes and cross--sections, and study the
hadronization rates for all these processes, which in turn are
necessary elements for constructing a dynamical non--equilibrium
transport theory for this model Lagrangian\cite{zhawi}.  This should
aid us in understanding the role that chiral symmetry plays in
dynamical processes.  Our approach follows in part, and develops
further the calculation performed in $SU(2)$ in the chiral limit
described in the letter~\cite{su2had}.

In our study of the $SU(3)$ sector, we classify hadronization processes
according to the incoming quark and antiquark, finding a total of four
independent classes under the additional assumption of $SU(2)$ isospin
symmetry, where $m_u=m_d$.  We thus consider the processes with $u\bar
d$, $u\bar s$, $u\bar u$ and $s\bar s$ as incoming pairs, which we also
list. The mesons $M$ and $M'$ considered in the process are $\pi$, $K$
and $\eta$, which are the stable ones with respect to the strong
interaction.  Feynman diagrams for the scattering amplitudes pertaining
to these reactions have a generic form which describes all $q\bar q\to
MM'$. We select the diagrams according to an expansion in the inverse
number of colors in the model, $1/N_c$, that is commensurate with the
Hartree approximation for the self--energy and the random phase
approximation for the scattering amplitude\cite{quack,mueller,DSTL}.
The variations in applying each of the Feynman diagrams to the
different scattering channels arise from differing exchanged mesons,
quark masses and flavor factors. We choose to illustrate our
calculation by analytically constructing the formulae for the cross
sections of the processes $u\bar d\to\pi^+\pi^0$, $u\bar
s\to\pi^+\overline{K^0}$ and $u\bar u\to\pi^+\pi^-$. The first of these
explicitly includes a $u$ channel exchange, while the second has
unequal masses for all incoming and outgoing particles. The final
process illustrates the role of the mixing of the scalar mesons in the
$SU(3)$ Lagrangian due to the t'Hooft term and the differing up, down
and strange quark masses.

All quantities are calculated as a function of temperature and baryonic
chemical potential. It turns out that a cardinal role is played by the
Mott temperatures, $T_{M\pi}$, $T_{M\sigma}$, $T_{MK}$, $T_{M\eta}$,
that are defined to be the temperatures at which the masses of the
respective mesons $\pi$, $\sigma$, $K$, $\eta$, are equal to the sum of
the masses of their constituents. Note that these four mesons are bound
states at $T=0$. At the Mott temperatures, the respective meson--quark
couplings go to zero, and this in turn influences the transition
amplitudes and transition rates in that they are suppressed. In
particular the pionic Mott temperature plays the dominant role, driving
the total transition rates (almost) to zero at this point. We comment
that the use of a finite temperature explicitly breaks Lorentz
invariance in our calculation. We do not attempt to resolve this
problem here. Quantities are simply calculated with respect to the rest
frame of the medium.

In order to make a connection with transport theory, we calculate
transition rates, which are constructed by multiplying the total cross
section of a particular process with the relative velocity.  Our
numerical results indicate that in a per--channel calculation, the $t$
and $u$ channel exchange of any one process dominate over the $s$
channel exchange, except perhaps in an energy range where a resonance
is present.  The $t$ and $u$ channel graphs, if present in a process,
give cancelling contributions at threshold when the flavor factors are
equal. In dealing with the four independent hadronization classes, we
are directly able to identify the leading contributions at all
energies.  These, with the exception of the $s\bar s$ hadronization
class, are dominated by pion production. The $s\bar s$ processes on the
other hand are found to be dominated by kaon production. In all these
calculations we can confirm that the exothermic reactions are always
dominant over the endothermic ones, as far as they are present.

We also examine the energy averaged transition rates. These quantities,
multiplied by the density of incoming quarks, can be interpreted as the
inverse of the hadronization time.  We find this to be of the order of
$\tau=2\mbox{--}4\mbox\,{\rm fm}/c$ in the temperature range
160--200\,MeV, rising rapidly as one moves towards the pion Mott
temperature $T_{M\pi}$, or to lower temperatures. This indicates that
hadronization occurs preferentially over this given range of
temperatures within our model.

Finally we have also examined processes that change the number of $s$
plus $\bar s$ quarks, like e.~g. $u\bar d\to K^+\overline{K^0}$.  We
find that the hadronization process generates an enhancement of 1\% in
the absolute value of the ratio of kaons to pions at $T=150\,{\rm
MeV}$.  This is to be compared with the experimental value of $(15.4\pm
0.8)\%$ for the $K/\pi$ ratio found for $S+S$ collisions, which itself
constitutes an enhancement of 9\% over the $N+N$ value of $(6.5\pm
1.1)\%$ \cite{na35}.

In order to facilitate the technical evaluation, we decompose all
quantities into fundamental integrals containing one, two or three
denominators, and which arise for instance from the self--energy,
polarization and the three meson vertex. In principle, these functions
can be calculated for arbitrary values of the differing quark masses
and associated differing chemical potentials. In this way, general
expressions can be constructed if one utilizes the modular forms.  For
actual calculational purposes, however, we will set all chemical
potentials to zero, and study temperature and energy dependencies here,
giving some relevant analytical results for these integrals in an
appendix.

This paper is structured as follows: In Section~\ref{mesonsec}, we
discuss the general $SU_L(3)\times SU_R(3)$ Lagrangian that we use, and
give the relevant functions associated with the mass spectrum
evaluation that we require. In particular the $\eta$--$\eta'$ mixing is
detailed, and the scalar resonance sector, which plays the role of
intermediate states in the hadronization.  In Section~\ref{hadrosec},
the possible hadronization processes are classified, and the explicit
examples $u\bar d\to \pi^+\pi^0$, $u\bar s\to \pi^+K^0$ and $u\bar
u\to\pi^+\pi^-$ are discussed. Numerical results for the transition
rates for each hadronization class are given in
Section~\ref{resultsec}. Strangeness changing processes are also
analyzed here.  We summarize and conclude in Section~\ref{sumsec}.

\section{Properties of Mesons in $SU(3)$}
\label{mesonsec}
\subsection{General Considerations}
This section serves to introduce our Lagrange density and notation,
commencing with the three flavor Lagrangian
\begin{eqnarray}
{\cal L} &=& \sum_{f=0}^3 \bar\psi_f(i\partial\! \! \! /-m_{0f})\psi_f
	\nonumber \\ & & +
	G\sum_{a=0}^8\left[(\bar\psi\lambda^a\psi)^2 +
			  (\bar\psi i\gamma_5\lambda^a\psi)^2\right]
	\nonumber \\ & & -
	K\left[\det\bar\psi(1+\gamma_5)\psi + \det\bar\psi(1-\gamma_5)\psi
				\right]
\label{lagrange1} \quad,
\end{eqnarray}
where $G$ and $K$ are dimensionful coupling strengths. The term
containing $G$ displays $U(3)\times U(3)$ symmetry, while the
determinantal term controlled by $K$ breaks this down to $SU(3)\times
SU(3)$. Flavor and color indices have been suppressed for convenience in
the interaction terms. However, the flavor indices have been {\em
explicitly\/} included in the first term, since the current quark masses
$m_{0f}$, that themselves explicitly break the $SU(3)\times SU(3)$
symmetry, are regarded as distinct. For general reviews on the three
flavor version of the NJL model, the reader is referred to
Refs.~\cite{vogl,spk,hatku}.

It is useful to convert the determinantal term in Eq.~(\ref{lagrange1})
into an effective two--body term in the mean field approximation. This
follows on contracting out one pair of quark and antiquark fields and
dividing this result by two. One may recombine the result with the
existing two body term that is controlled by $G$ to find the effective
Lagrangian to be
\begin{eqnarray}
{\cal L} &=& \sum_{f=0}^3 \bar\psi_f(i\partial\! \! \! /-m_{0f})\psi_f
        \nonumber \\ & & +
	\sum_{a=0}^8\left[K_a^-(\bar\psi\lambda^a\psi)^2 +
			K_a^+(\bar\psi i\gamma_5\lambda^a\psi)^2\right]
        \nonumber \\ & & +
	K_{30}^-(\bar\psi\lambda^3\psi)(\bar\psi\lambda^0\psi) +
	K_{30}^+(\bar\psi i\gamma_5\lambda^3\psi)
		(\bar\psi i\gamma_5\lambda^0\psi)
        \nonumber \\ & & +
	K_{03}^-(\bar\psi\lambda^0\psi)(\bar\psi\lambda^3\psi) +
	K_{03}^+(\bar\psi i\gamma_5\lambda^0\psi)
		(\bar\psi i\gamma_5\lambda^3\psi)
        \nonumber \\ & & +
	K_{80}^-(\bar\psi\lambda^8\psi)(\bar\psi\lambda^0\psi) +
	K_{80}^+(\bar\psi i\gamma_5\lambda^8\psi)
		(\bar\psi i\gamma_5\lambda^0\psi)
        \nonumber \\ & & +
	K_{08}^-(\bar\psi\lambda^0\psi)(\bar\psi\lambda^8\psi) +
	K_{08}^+(\bar\psi i\gamma_5\lambda^0\psi)
		(\bar\psi i\gamma_5\lambda^8\psi)
        \nonumber \\ & & +
	K_{83}^-(\bar\psi\lambda^8\psi)(\bar\psi\lambda^3\psi) +
	K_{83}^+(\bar\psi i\gamma_5\lambda^8\psi)
		(\bar\psi i\gamma_5\lambda^3\psi)
        \nonumber \\ & & +
	K_{38}^-(\bar\psi\lambda^3\psi)(\bar\psi\lambda^8\psi) +
	K_{38}^+(\bar\psi i\gamma_5\lambda^3\psi)
		(\bar\psi i\gamma_5\lambda^8\psi)
\label{lagrange2}
\end{eqnarray}
with the effective coupling constants
\begin{mathletters}
\label{cpleff}
\begin{equation}
K_0^\pm = G \mp \frac{1}{3} N_c K(i{\rm tr}_\gamma S^u(x,x)+
i{\rm tr}_\gamma S^d(x,x)+i{\rm tr}_\gamma S^s(x,x))
\end{equation} \begin{equation}
K_1^\pm=K_2^\pm=K_3^\pm = G\pm\frac{1}{2}N_cKi{\rm tr}_\gamma S^s(x,x)
\end{equation} \begin{equation}
K_4^\pm=K_5^\pm = G\pm\frac{1}{2}N_cKi{\rm tr}_\gamma S^d(x,x)
\end{equation} \begin{equation}
K_6^\pm=K_7^\pm = G\pm\frac{1}{2}N_cKi{\rm tr}_\gamma S^u(x,x)
\end{equation} \begin{equation}
K_8^\pm = G\pm\frac{1}{6}N_cK(2i{\rm tr}_\gamma S^u(x,x)+
2i{\rm tr}_\gamma S^d(x,x) - {\rm tr}_\gamma S^s(x,x))
\end{equation} \begin{equation}
K_{03}^\pm=K_{30}^\pm = \mp \frac{1}{2\sqrt{6}}
N_c K(i{\rm tr}_\gamma S^u(x,x)
- i{\rm tr}_\gamma S^d(x,x))
\end{equation} \begin{equation}
K_{08}^\pm=K_{80}^\pm = \pm \frac{\sqrt{2}}{12} N_c
K(i{\rm tr}_\gamma S^u(x,x) + i{\rm tr}_\gamma S^d(x,x)
- 2i{\rm tr}_\gamma S^s(x,x))
\end{equation} \begin{equation}
K_{38}^\pm=K_{83}^\pm = \pm \frac{1}{2\sqrt{3}} N_c
K(i{\rm tr}_\gamma S^u(x,x) - i{\rm tr}_\gamma S^d(x,x))
\quad.
\end{equation}
\end{mathletters}
In Eq.~(\ref{cpleff}), ${\rm tr}_\gamma$ refers to the spinor
trace alone, while $S^f$ is the diagonal quark propagator for a given
flavor $f$, which, in the imaginary time formalism for finite
temperatures can be written as
\begin{equation}
S^f(\vec x - \vec x', \tau - \tau') = \frac{i}{\beta}\sum_n
e^{-i\omega_n(\tau-\tau')} \int \frac{d^3p}{(2\pi)^3}
\frac{e^{i\vec p(\vec x-\vec x')}}
{\gamma_0(i\omega_n+\mu_f) - \vec\gamma \vec p -m_f} \quad.
\end{equation}
Here the Matsubara frequencies are fermionic, $i\omega_n=(2n+1)\pi/\beta$,
with $n=0,\pm 1,\pm 2,\pm 3,\dots$, and $\mu_f$ is the chemical
potential for a quark of flavor $f$.

As can be seen from Eqs.~(\ref{lagrange2}, \ref{cpleff}), the
nondiagonal coupling constants $K_{03}^+$, $K_{08}^+$ and $K_{38}^+$ give
rise to the $\pi^0$--$\eta$--$\eta'$ mixing.  If one assumes $SU(2)$
isospin symmetry, i.~e. $m_u=m_d$, then $K_{03}^+$ and $K_{38}^+$ vanish
identically, with the consequence that the $\pi^0$ decouples from the
$\eta$ and $\eta'$. We make this assumption in what follows, using the
generic label $q$ for $u$, $d$, and always writing $s$ explicitly.

Writing
\begin{equation}
i{\rm tr}_\gamma S^f(x,x) = - \frac{m_f}{4\pi^2}A(m_f, \mu_f) \quad,
\end{equation}
where
\begin{equation}
A(m_f, \mu_f) =
\frac{16\pi^2}{\beta} \sum_n e^{i\omega_n\eta}
\int\limits_{|\vec p| < \Lambda} \frac{d^3p}{(2\pi)^3}
\frac{1}{(i\omega_n+\mu_f)^2-E_f^2}
\label{Adef}
\end{equation}
(with $E_f^2=p^2+m_f^2$) denotes the first loop integral (cf.
Appendix~\ref{app-loop}), one can easily derive the coupled
gap-equations
\begin{mathletters}
\begin{equation}
m_u = m_{0u} - \frac{GN_c}{\pi^2}m_uA(m_u, \mu_u)
             + \frac{KN_c^2}{8\pi^4}m_dm_sA(m_d, \mu_d) A(m_s, \mu_s)
\end{equation} \begin{equation}
m_d = m_{0d} - \frac{GN_c}{\pi^2}m_dA(m_d, \mu_d)
             + \frac{KN_c^2}{8\pi^4}m_sm_uA(m_s, \mu_s) A(m_u, \mu_u)
\end{equation} \begin{equation}
m_s = m_{0s} - \frac{GN_c}{\pi^2}m_sA(m_s, \mu_s)
             + \frac{KN_c^2}{8\pi^4}m_um_dA(m_u, \mu_u) A(m_d, \mu_d)
\end{equation}
\end{mathletters}
{}from the mean field or Hartree approximation to the self--energy, and
which determine the physical quark masses.

\subsection{Pions and Kaons}
\label{pisect}
Imposing the degeneracy condition $m_u=m_d$, the determination of both
the pion and kaon masses follows similarly to the standard approach
taken in the two flavor model\cite{spk}. This comes about, since the
off--diagonal coupling strengths $K^\pm_{03}$ and $K^\pm_{38}$ are
identically zero, and the $\pi^0$ is decoupled from the $\eta$ and
$\eta'$. In this limit, the $\pi^\pm$, $\pi^0$ become degenerate.
Concomitantly, $K_4^\pm=K_6^\pm$, which means physically that neutral
and charged kaons have the same mass.

The quark--quark scattering amplitude is calculated in the random
phase approximation (RPA, see Fig.~\ref{scatgraph}), which yields the
result
\begin{equation}
M_\pi(k_0,\vec k) = \frac{2K^+_1}{1-4K^+_1\Pi_{q\bar q}^P(k_0, \vec k)}
\label{piprop}
\end{equation}
for the pion scattering amplitude and
\begin{equation}
M_K(k_0,\vec k) = \frac{2K^+_4}{1-4K^+_4\Pi_{q\bar s}^P(k_0, \vec k)}
\label{kaprop}
\end{equation}
for the kaon one. Here $\Pi^P(k_0, \vec k)$ represents the irreducible
pseudoscalar polarization (see Fig.~\ref{polgraph}), which, in the
finite temperature Matsubara formalism, depends on $k_0$ and $\vec k$
separately, because the medium breaks Lorentz invariance.  For
arbitrary flavors, the irreducible polarization is given by the analytic
continuation of the imaginary time form
\begin{eqnarray}
& &-i\Pi^P_{f_1f_2}(i\nu_m,\vec k)
= -N_c \frac{i}{\beta}\sum_n \int \frac{d^3p}{(2\pi)^3}
{\rm tr}_\gamma \left[ iS^{f_1}(i\omega_n,\vec p)
i\gamma_5 iS^{f_2}(i\omega_n-i\nu_m,\vec p-\vec k) i\gamma_5 \right]
\nonumber \\ & & \quad
= 4iN_c \frac{1}{\beta}\sum_n \int \frac{d^3p}{(2\pi)^3}
\frac{(i\omega_n + \mu_1)(i\omega_n-i\nu_m+\mu_2) - \vec p(\vec p - \vec k)
-m_1m_2}{\left[(i\omega_n+\mu_1)^2-E_1^2\right]
\left[(i\omega_n-i\nu_m+\mu_2)^2 -E_2^2\right] }
\end{eqnarray}
where $E_1 = \sqrt{\vec p^2+m_1^2}$, $E_2 = \sqrt{(\vec p - \vec
k)^2-m_2^2}$, and the mesonic Matsubara frequencies $i\nu_m=2m\pi/\beta$
are even, $m=0,\pm 1,\pm 2, \dots$.  Note that the irreducible
polarization in this definition does not contain any flavor factors from
the Gell--Mann $\lambda$-matrices.  These factors are incorporated
explicitly in Eqs.~(\ref{piprop}, \ref{kaprop}) as multiplicative
cofactors of the coupling and polarization.

We may decompose $\Pi^P(k_0,\vec k)$ in terms of the function $A(m, \mu)$
already defined in Eq.~(\ref{Adef}) as
\begin{eqnarray}
\Pi^P(k_0,\vec k) &=& -\frac{N_c}{8\pi^2}
\Big[A(m_1, \mu_1)+A(m_2, \mu_2)
\label{polar} \\ & & +
\left((m_1-m_2)^2-(k_0+\mu_1-\mu_2)^2+\vec k ^2\right)
B_0(\vec k, m_1, \mu_1, m_2, \mu_2, k_0) \Big]
\nonumber \quad,
\end{eqnarray}
introducing the second loop integral (cf. Appendix~\ref{app-loop})
\begin{eqnarray}
& &B_0(\vec k, m_1, \mu_1, m_2, \mu_2, i\nu_m) =
\label{b0def} \\ & &
\frac{16\pi^2}{\beta} \sum_n e^{i\omega_n\eta}
\int\limits_{|\vec p| < \Lambda} \frac{d^3p}{(2\pi)^3}
\frac{1}{((i\omega_n+\mu_1)^2-E_1^2)}
\frac{1}{((i\omega_n-i\nu_m+\mu_2)^2-E_2^2)}
\nonumber \quad.
\end{eqnarray}
(Note that in Eq.~(\ref{polar}), an analytic continuation has been
performed.) We comment that, due to rotational invariance, $B_0$ does
not fully depend on $\vec k$, but only on $|\vec k|$.

In terms of the modular integrals, $A$ and $B_0$, polarizations
required can then be explicitly given for arbitrary chemical potential
and temperature as
\begin{equation}
\Pi_{q\bar q}^P(k_0, \vec k) = -\frac{N_c}{8\pi^2}
\left[2A(m_q, \mu_q)-k^2B_0(\vec k, m_q, \mu_q, m_q, \mu_q, k_0)\right]
\label{piqq}
\end{equation} \begin{equation}
\Pi_{s\bar s}^P(k_0, \vec k) = -\frac{N_c}{8\pi^2}
\left[2A(m_s, \mu_s)-k^2B_0(\vec k, m_s, \mu_s, m_s, \mu_s, k_0)\right]
\label{piss}
\end{equation} \begin{eqnarray}
\Pi_{q\bar s}^P(k_0, \vec k) &=& -\frac{N_c}{8\pi^2}
\Big[A(m_q, \mu_q)+A(m_s, \mu_s)
\\ & & +
\left((m_q-m_s)^2-(k_0+\mu_q-\mu_s)^2+\vec k ^2\right)
B_0(\vec k, m_q, \mu_q, m_s, \mu_s, k_0) \Big]
\quad. \nonumber
\end{eqnarray}

The pion and kaon masses are determined according to the dispersion
relations\cite{spk}
\begin{equation}
1-4K^+_1\Pi_{q\bar q}^P (m_\pi, \vec 0) = 0 \label{pimass}
\end{equation} \begin{equation}
1-4K^+_4\Pi_{q\bar s}^P (m_K, \vec 0) = 0 \label{kmass}
\quad,
\end{equation}
while effective couplings can be identified from the pole
approximation forms
\begin{equation}
M_\pi(k_0,\vec k) \approx \frac{-g_{\pi q \bar q}^2}{k^2-m_\pi^2}
\end{equation} \begin{equation}
M_K(k_0,\vec k) \approx \frac{-g_{K q \bar s}^2}{k^2-m_K^2}
\end{equation}
to be
\begin{equation}
g_{\pi q\bar q}^{-2} = - \frac{1}{2m_\pi}
\left. \frac{\partial\Pi_{q\bar q}^P(k_0, \vec 0)}{\partial k_0}
\right|_{k_0=m_\pi} \label{cplpi}
\end{equation} \begin{equation}
g_{K q\bar s}^{-2} = - \frac{1}{2m_K}
\left. \frac{\partial\Pi_{q\bar s}^P(k_0, \vec 0)}{\partial k_0}
\right|_{k_0=m_K} \label{cplk}
\end{equation}
At low temperatures, Eqs.~(\ref{pimass}, \ref{kmass}) have bound state
solutions with $m_\pi < 2m_q$ and $m_K < m_q + m_s$. In this case,
Eqs.~(\ref{pimass}, \ref{kmass}) are real equations. At higher
temperatures, the polarization functions become complex functions
with complex solutions for the meson masses, that we may write
as\cite{njlthermo}
\begin{equation}
m_\pi \to m_\pi - \frac{i}{2}\Gamma_\pi
\end{equation} \begin{equation}
m_K \to m_K - \frac{i}{2}\Gamma_K \quad.
\end{equation}
Denoting the
Mott transition temperatures $T_{M\pi}$ and $T_{MK}$ as the
temperatures at which $m_\pi=2m_q$ and $m_K=m_q+m_s$ respectively, one
thus has the physical circumstance that at temperatures larger than
$T_{M\pi}$ or $T_{MK}$, respectively pions or kaons become resonances
with finite widths due to the available decay channels into two quarks.
{}From Eqs.~(\ref{cplpi}, \ref{cplk}), it follows that the quark--meson
couplings also become complex in this case.

\subsection{$\eta$ and $\eta'$}
\label{etasect}
Because of the mixing terms occurring in Eq.~(\ref{lagrange2}), the
calculation of the $\eta$ and $\eta'$ masses and couplings is somewhat
more involved.  The scattering amplitude is nondiagonal in this sector
with entries $M_{00}$, $M_{88}$, and $M_{08} = M_{80}$. Within the RPA,
that is still expressed diagramatically in Fig.~\ref{scatgraph}, it can
be calculated in matrix form to be\cite{spk}
\begin{equation}
M = 2K^+(1-2\Pi^PK^+)^{-1} \label{mixprop}
\quad,
\end{equation}
where now $K^+$ and $\Pi^P$ are the $2\times 2$ matrices
\begin{equation}
K^+ = \left( \begin{array}{cc} K^+_0 & K^+_{08} \\
       K^+_{80} & K^+_8 \end{array} \right) \label{eq36}
\end{equation} \begin{equation}
\Pi^P = \left( \begin{array}{cc} \Pi^P_0 & \Pi^P_{08} \\ \Pi^P_{80} &
\Pi^P_8 \end{array} \right) \label{eq37} \quad.
\end{equation}
The $K^+_i$ on the right hand side of Eq.~(\ref{eq36}) are as defined
in Eqs.~(\ref{cpleff}), while the pseudoscalar polarization functions
are the linear combinations
\begin{mathletters}
\label{eq40}
\begin{equation}
\Pi_0^P = \frac{2}{3}\left(2\Pi_{q\bar q}^P+\Pi_{s\bar s}^P\right)
\end{equation} \begin{equation}
\Pi_8^P = \frac{2}{3}\left(\Pi_{q\bar q}^P+2\Pi_{s\bar s}^P\right)
\end{equation} \begin{equation}
\Pi_{08}^P = \Pi_{80}^P = \frac{2\sqrt{2}}{3}\left(\Pi_{q\bar q}^P-
\Pi_{s\bar s}^P\right)
\end{equation}
\end{mathletters}
of the functions $\Pi^P_{q\bar q}$ and $\Pi^P_{s\bar s}$, that can be
evaluated via Eqs.~(\ref{piqq}, \ref{piss}).  In Eqs.~(\ref{mixprop}) to
(\ref{eq40}), we have dropped the argument $(k_0,\vec k)$ for
convenience.  A determination of the masses and coupling strengths can
be made on forming the inverse of the matrix $M$. We abbreviate this as
\begin{equation}
M^{-1}=\frac{1}{2{\rm det\/}K^+}
\left( \begin{array}{cc} {\cal A} & {\cal B} \\
\cal{B} & \cal{C} \end{array} \right)
\label{eq41}
\end{equation}
with
\begin{mathletters}
\begin{equation}
{\cal A} = K^+_8 -2 \Pi_0^P{\rm det\/}K^+ \\
\end{equation} \begin{equation}
-{\cal B} = K^+_{08} +2 \Pi_{08}^P{\rm det\/}K^+ \\
\end{equation} \begin{equation}
{\cal C} = K^+_0 -2 \Pi_8^P{\rm det\/}K^+
\quad.
\end{equation}
\end{mathletters}
Following \cite{hatku}, we introduce the diagonal forms $M_\eta$ and
$M_{\eta'}$ via
\begin{equation}
M^{-1}=\frac{1}{4{\rm det\/}K^+}
\left(\begin{array}{cc} c & -s \\ s & c \end{array}\right)
\left(\begin{array}{cc} M_\eta^{-1} & 0 \\
0 & M_{\eta'}^{-1} \end{array}\right)
\left(\begin{array}{cc} c & s \\ -s & c \end{array}\right)
\quad,
\end{equation}
where it is a simple matter to verify that
\begin{mathletters}
\begin{equation}
M_\eta^{-1} = {\cal A} + {\cal C} - \sqrt{({\cal A}-{\cal C})^2 +
		4{\cal B}^2}
\end{equation} \begin{equation}
M_{\eta'}^{-1} = {\cal A} + {\cal C} + \sqrt{({\cal A}-{\cal C})^2 +
		4{\cal B}^2}
\end{equation}
\end{mathletters} \begin{mathletters}
\begin{equation}
c^2 + s^2 = 1
\end{equation} \begin{equation}
c^2-s^2 = \frac{{\cal C - A}}{\sqrt{({\cal A}-{\cal C})^2 + 4{\cal B}^2}}
\end{equation} \begin{equation}
2cs = \frac{{-2 \cal B}}{\sqrt{({\cal A}-{\cal C})^2 + 4{\cal B}^2}}
\quad.
\end{equation}
\end{mathletters}
The masses of the $\eta$ and $\eta'$ mesons can now be determined via
the condition
\begin{equation}
M_\eta^{-1}(m_\eta, \vec 0) = 0 \label{etamass}
\end{equation} \begin{equation}
M_{\eta'}^{-1}(m_{\eta'}, \vec 0) = 0 \label{etapmass}
\quad.
\end{equation}
For meson masses that lie below the quark--antiquark mass threshold,
these equations are again real. In practice, actual numerical
determinations place the $\eta$ mass below this threshold at $T=0$, and
at a specific Mott transition temperature $T_{M\eta}$, this becomes a
resonance, in much the same way as occurs for the pions and kaons.  The
$\eta'$ meson, however, is distinguished by the fact that it {\em
always\/} lies above the quark--antiquark threshold, and it therefore is
a resonant state at all temperatures. Since this is an artifact due to
the lack of confinement, we regard this feature with some scepticism.
We nevertheless calculate its contribution to the relevant hadronization
cross sections and transition rates, and find that it is negligible, so
that it can be safely discarded.

To calculate the coupling constants from Eq.~(\ref{eq41}), we express
$M$ directly as
\begin{equation}
M = \frac{2}{{\cal D}} \left( \begin{array}{cc} {\cal C} & - {\cal B} \\
	- {\cal B} & {\cal A} \end{array} \right)
\end{equation}
with
\begin{eqnarray}
{\cal D} &=& ({\cal AC} - {\cal B}^2) / \det K^+ \nonumber \\
	&=& 1 - 2{\rm tr}(\Pi K^+) + 4\det \Pi \det K^+
\quad.
\end{eqnarray}
At $(k_0,\vec k)=(m_\eta,\vec 0)$, ${\cal D} = 0$, and in the usual
fashion, we make a pole
approximation giving the form
\begin{eqnarray}
M &=& \frac{4m_\eta}{\left. \frac{\partial D}{\partial k_0}
     \right|_{k_0=m_\eta}} \frac{1}{k^2-m_\eta^2}
\left( \begin{array}{cc} {\cal C} & - {\cal B} \\
	- {\cal B} & {\cal A} \end{array} \right)
\end{eqnarray}
{}From this, one obtains
\begin{equation}
M_{ab} = -\frac{g_{a\eta}g_{b\eta}}{k^2-m_\eta^2}
\end{equation}
with
\begin{equation}
g_{0\eta}^2 = -\frac{4m_\eta{\cal C}}
{\left. \frac{\partial D}{\partial k_0}
\right|_{k_0=m_\eta}}
\end{equation} \begin{equation}
g_{8\eta}^2 = -\frac{4m_\eta{\cal A}}
{\left. \frac{\partial D}{\partial k_0}
\right|_{k_0=m_\eta}}
\end{equation} \begin{equation}
g_{0\eta}g_{8\eta} = \frac{4m_\eta{\cal B}}{\left.
          \frac{\partial D}{\partial k_0}
\right|_{k_0=m_\eta}} \quad.
\end{equation}
{}From these coupling constants we may calculate \cite{hatku}
\begin{equation}
g_{\eta q \bar q} =
	\sqrt{\frac{2}{3}} g_{0\eta} + \frac{1}{\sqrt{3}}g_{8\eta}
\end{equation} \begin{equation}
g_{\eta s \bar s} =
	\sqrt{\frac{2}{3}} g_{0\eta} - \frac{2}{\sqrt{3}}g_{8\eta}
\quad.
\end{equation}
The coupling strengths $g_{\eta'q\bar q}$ and $g_{\eta's\bar s}$ may be
evaluated in the same fashion. The set of couplings $g_{\eta q\bar q}$,
$g_{\eta s\bar s}$ and $g_{\eta'q\bar q}$, $g_{\eta's\bar s}$ enter
directly into the cross section calculations.

\subsection{Scalar Resonances}
Since scalar mesons may occur as possible intermediate resonance
structures in the hadronization cross sections, we are also interested
in obtaining their masses. In contrast to the two flavor model, in
which only the $\sigma$ is present, we deal here with nine scalar
resonances:  three $\sigma_\pi$'s, which are the scalar partners of the
pions, four $\sigma_K$'s, being the scalar partners of the kaons, and
the $\sigma$ and $\sigma'$, that are associated similarly with the
$\eta$ and $\eta'$.  As occurs in the pseudoscalar case, for $m_u\ne
m_d\ne m_s$ we have mixing between the $\sigma$, $\sigma'$ and the
neutral $\sigma_\pi$, the latter decoupling from the former if $SU(2)$
isospin degeneracy, $m_u=m_d$, is imposed.

The same techniques applied in Secs.~\ref{pisect} and \ref{etasect} can
now be directly applied to the scalar resonances, with two changes: (i)
one has to replace the coupling constants $K^+_i$ of the previous
section by $K^-_i$ given in Eq.~(\ref{cpleff}) and (ii) the
pseudoscalar polarizations are replaced by their scalar counterparts:
\begin{equation}
\Pi_{q\bar q}^S(k_0, \vec k) = -\frac{N_c}{8\pi^2}
\left[2A(m_q, \mu_q)+(4m_q^2-k^2)B_0(\vec k, m_q, \mu_q, m_q, \mu_q,
k_0)\right]
\end{equation} \begin{equation}
\Pi_{s\bar s}^S(k_0, \vec k) = -\frac{N_c}{8\pi^2}
\left[2A(m_s, \mu_s)+(4m_s^2-k^2)B_0(\vec k, m_s, \mu_s, m_s, \mu_s,
k_0)\right]
\end{equation} \begin{eqnarray}
\Pi_{q\bar s}^S(k_0, \vec k) &=& -\frac{N_c}{8\pi^2}
\Big[A(m_q, \mu_q)+A(m_s, \mu_s)
\\ & & +
\left((m_q+m_s)^2-(k_0+\mu_q-\mu_s)^2+\vec k ^2\right)
B_0(\vec k, m_q, \mu_q, m_s, \mu_s, k_0) \Big] \quad,
\nonumber
\end{eqnarray}
which are derived from the same graph as in Fig.~\ref{polgraph}, dropping
the $i\gamma_5$ factors at the vertices.

\subsection{Numerical Results}
For our numerical calculations, we employ the parameter set $m_{0q} =
5.5\,{\rm MeV}$, $m_{0s} = 140.7\,{\rm MeV}$, $G\Lambda^2 = 1.835$,
$K\Lambda^5 = 12.36$ and $\Lambda = 602.3\,{\rm MeV}$, that has been
determined on fixing the conditions $m_\pi=135.0\,{\rm MeV}$,
$m_K=497.7\,{\rm MeV}$, $m_{\eta'}=957.8\,{\rm MeV}$ and
$f_\pi=92.4\,{\rm MeV}$, while $m_{0q}$ is fixed at 5.5\,MeV.  The
reason why we fitted the mass of the $\eta$ instead of the $\eta'$ has a
purely technical origin. This parameter set gives an $\eta$ mass of
$m_\eta=514.8\,{\rm MeV}$, which compares reasonably well with the
physical value $m_\eta=548.8\,{\rm MeV}$.  Although we have developed
the general formalism to include the case of finite chemical potentials,
we confine ourselves to $\mu_q=\mu_s=0$ in what follows.

Figure~\ref{gap} shows the temperature dependence of the constituent
quark masses. At $T=0$, we find $m_q=367.7\,{\rm MeV}$ and
$m_s=549.5\,{\rm MeV}$.  At temperatures around 200\,MeV the mass of the
light quarks drops to the current quark mass, indicating a washed out
crossover from the chirally broken to approximately chirally symmetric
phase. The strange quark mass starts to decrease significantly in this
temperature range, but even at $T=300\,{\rm MeV}$ it is still a factor
of two away from the strange current quark mass.

Figure \ref{massplot} shows the temperature dependence of the
pseudoscalar meson masses at $\mu=0$. For comparison, the curves of
$2m_q$ and $m_q+m_s$ are also indicated. At low temperatures, the meson
masses are approximately constant. The crossing of the $\pi$ and $\eta$
lines with the $2m_q$ line indicates the respective Mott transition
temperature for these particles, $T_{M\pi}$ and $T_{M\eta}$.  One
observes that $T_{M\eta}<T_{M\pi}$, the absolute values are
$T_{M\eta}=180\,{\rm MeV}$ and $T_{M\pi}=212\,{\rm MeV}$.  For
temperatures higher than $T_{M\pi}$, $T_{M\eta}$, respectively, the
$\pi$ and $\eta$ become resonances and their masses increase.  Similarly
a Mott transition temperature $T_{MK}$ for the kaon modes can be defined
at the point where $m_K$ meets $m_s+m_q$. This is also indicated in the
figure. One can also see from this plot that $T_{M\pi}$ and $T_{MK}$ are
approximately equal, with $T_{MK}=210\,{\rm MeV}$.

Figure \ref{cplplot} shows the absolute values of the pion and kaon
coupling constants. A striking behaviour is observed at the Mott
temperature. There the polarization displays a kink singularity, which
can also be seen in the meson masses.  Technically this results in the
coupling strengths approaching zero for $T\to T_M$ from below. This
behaviour differs markedly from the behaviour of the couplings when
evaluated in the chiral limit. In that case, the coupling strength is
{\em always\/} different from zero as one approaches the transition
temperature from below.  Note that this strong deviation from the chiral
limit behaviour may have extreme consequences for results that depend
strongly on this function, such as the cross sections. Since we regard
the physical situation as being non--chiral, we investigate this
situation only, and draw conclusions accordingly.  The $\eta$--quark
couplings have not been explicitly shown; they display a qualitatively
similar behaviour.

\begin{figure} % Figure 6
\caption[]{Temperature dependence of the scalar meson masses and $2m_q$.}
      \label{scalarplot}
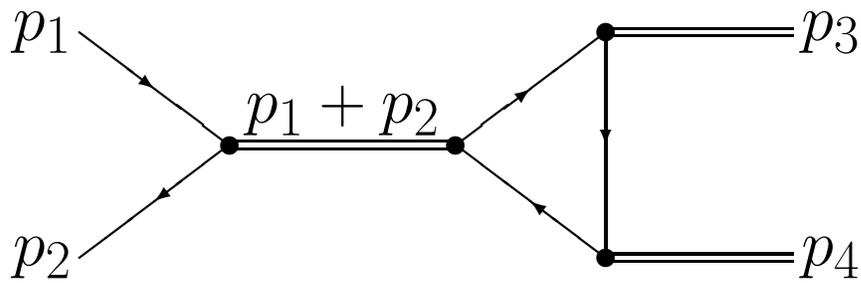
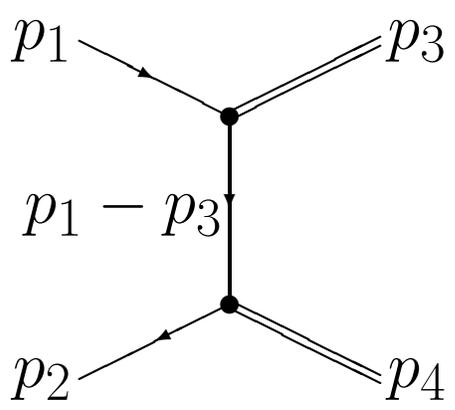
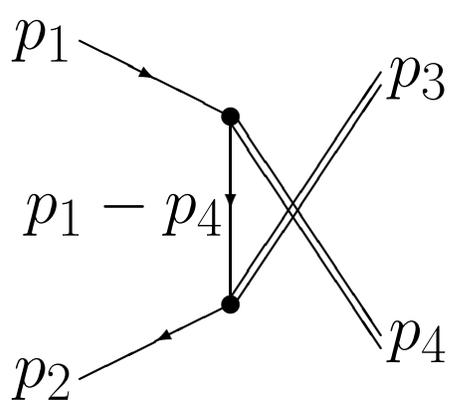
\end{figure}

The numerical calculations of the scalar mass spectrum are shown in
Fig.~\ref{scalarplot}. At $T=0$, we find $m_\sigma=728.9\,{\rm MeV}$,
$m_{\sigma'}=1198.3\,{\rm MeV}$, $m_{\sigma_\pi}=880.2\,{\rm MeV}$ and
$m_{\sigma_K}=1050.5\,{\rm MeV}$. For comparison, the double constituent
quark mass is also shown in the figure. All these mesons are unstable
over the entire temperature range, except for the $\sigma$, for which
the mixing results in a mass slightly below $2m_q$ at $T=0$. Although the
difference is small compared to the standard two flavor model, it has
the qualitative effect of making the $\sigma$ a stable particle for
temperatures up to its Mott temperature $T_{M\sigma}=165\,{\rm MeV}$.
As in the two flavor model, we obtain $m_\sigma\approx m_\pi$ above the
pion Mott temperature, as is expected from symmetry requirements.

The results presented in this section compare well with those of
Ref.~\cite{hatku}.

\section{Hadronization Cross Sections}
\label{hadrosec}
\subsection{Classification of Hadronization Processes}
Since a large number of hadronization processes are available to the
light and strange sector quarks, it is useful to introduce a
classification scheme to simplify the task of bookkeeping. The cross
sections are classified according to the incoming quarks and include
all exit channels\cite{su2had}. Since we work in the approximation
$m_u=m_d$, we have isospin symmetry and charge conjugation, leading
to the relations
\begin{mathletters}
\begin{equation}
\sigma_{u\bar d} = \sigma_{d\bar u}
\end{equation} \begin{equation}
\sigma_{u\bar s} = \sigma_{d\bar s} = \sigma_{s\bar u} = \sigma_{s\bar d}
\end{equation} \begin{equation}
\sigma_{u\bar u} = \sigma_{d\bar d} \quad,
\end{equation}
\end{mathletters}
using an obvious notation. Together with $\sigma_{s\bar s}$, we thus
have four independent classes of hadronization cross sections,
$\sigma_{u\bar d}$, $\sigma_{u\bar s}$, $\sigma_{u\bar u}$ and
$\sigma_{s\bar s}$.  For these four classes, we determine the
hadronization processes that are not forbidden by charge, strangeness
or isospin conservation. These conservation laws lead to the processes
that are listed in Table~\ref{channeltab}.

For the outgoing channels in Table~\ref{channeltab}, further symmetry
relations hold:
\begin{mathletters}
\begin{equation}
\sigma_{u\bar s \to \pi^+ K^0} = 2 \sigma_{u\bar s \to \pi^0 K^+}
\label{sym1}
\end{equation} \begin{equation}
\sigma_{u\bar d \to \pi^+ \eta} = 2 \sigma_{u\bar u \to \pi^0 \eta}
\label{sym2}
\end{equation} \begin{equation}
\sigma_{u\bar d \to \pi^+ \eta'} = 2 \sigma_{u\bar u \to \pi^0 \eta'}
\label{sym3}
\end{equation} \begin{equation}
\sigma_{s\bar s \to K^+ K^-} = \sigma_{s\bar s \to K^0 \overline{K^0}}
\label{sym4}
\end{equation} \begin{equation}
\sigma_{s\bar s \to \pi^+ \pi^-} = 2 \sigma_{s\bar s \to \pi^0 \pi^0}
\label{sym5} \quad,
\end{equation}
\end{mathletters}
as a consequence of flavor algebra. In total, there are nineteen
independent cross sections that we calculate.

\subsection{Feynman Graphs}
\begin{figure} % Figure 7
\caption[]{Generic form of Feynman graphs for the hadronization
      amplitudes to leading order in $1/N_c$.  Quarks are denoted by
      single lines, mesons by double ones. The three diagrams represent
      $s$ channel, $t$ channel and $u$ channel exchanges, respectively.}
      \label{gengraph}
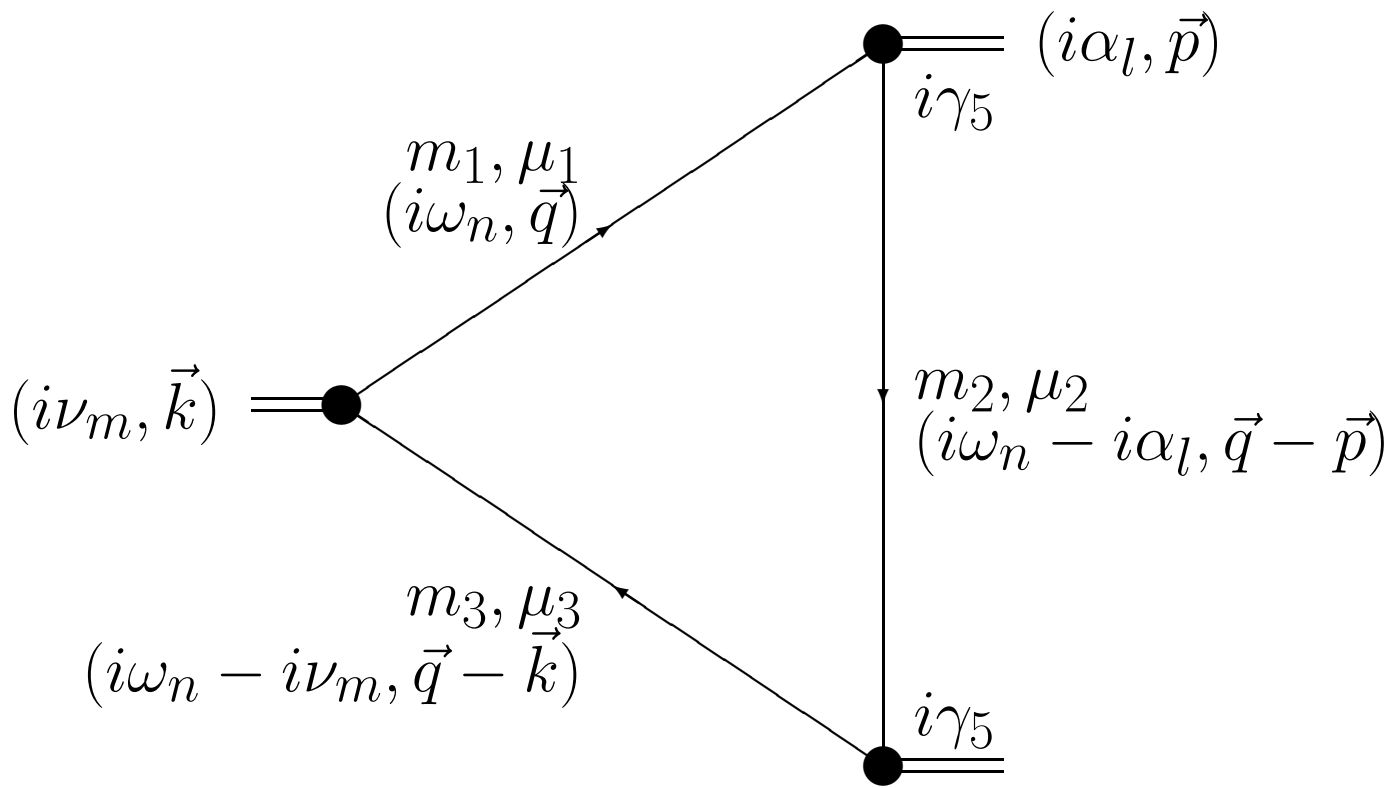
\end{figure}

The Feynman graphs that we consider, have the generic forms shown in
fig.~\ref{gengraph}. This choice of diagrams is in keeping with the
evaluation of the Hartree diagram to determine the gap equation and the
random phase approximation for the polarization. Together this selection
constitutes a consistent expansion in the inverse number of colors,
$1/N_c$\cite{quack,mueller,DSTL}. We stress that this is not an expansion
in the coupling strength, but is rather a non--perturbative expansion.
Note that the calculation of the transition amplitudes is complicated
with respect to the $SU(2)$ isospin symmetric case by the fact that here
each fermion line carries a flavor dependent mass, while the meson lines
in turn carry differing masses also. We therefore attempt to retain as
general a formalism as possible in what follows for the transition
amplitudes and we specify the parameters later.

\subsubsection{$s$ channel}
\begin{figure} % Figure 8
\caption[]{Three meson vertex, $\Gamma(i\nu_m,\vec k;i\alpha_l,\vec p)$.}
      \label{trigraph}
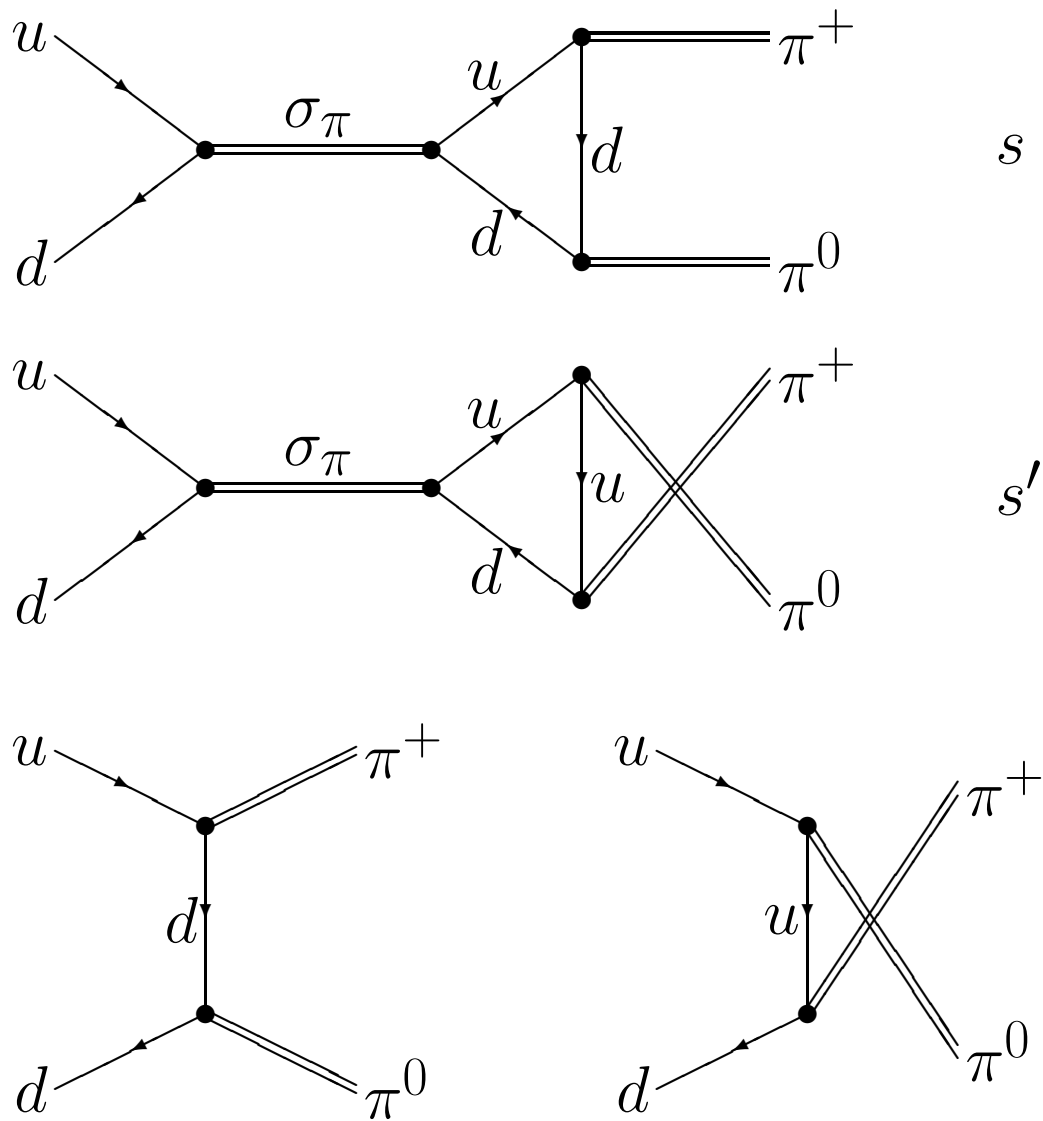
\end{figure}

The $s$ channel exchange diagrams have the form
\begin{equation}
-i {\cal M}_s = \bar v(p_2) u(p_1) \delta_{c_1c_2} f_s i{\cal D}(p_1+p_2)
		\Gamma(p_1+p_2; p_3) ig_1 ig_2 \label{eq72} \quad,
\end{equation}
where $g_1$, $g_2$ are meson--quark coupling strengths for the outgoing
mesons and $f_s$ is a flavor factor. The momenta of the
incoming particles are $p_1$ and $p_2$. For the outgoing
particles, we assign momenta $p_3$ and $p_4$.
$\cal D$ stands for the scattering amplitude of the virtual
scalar meson. It can either take the form appropriate for a scalar
meson that corresponds to Eqs.~(\ref{piprop}, \ref{kaprop}), if the
incoming quarks have different flavor, or it can be constructed as the
sum over mixing terms according to Eq.~(\ref{mixprop}), in all cases
with $K_i^+$ replaced by $K_i^-$, and $\Pi^P$ by $\Pi^S$. The symbol
$\Gamma$ describes the three meson vertex contribution to the diagram.
Its general form in the imaginary time formalism
is (cf. Fig.~\ref{trigraph})
\begin{eqnarray}
& &\Gamma(i\nu_m,\vec k; i\alpha_l, \vec p)
= -N_c \frac{i}{\beta} \sum_n \int \frac{d^3q}{(2\pi)^3}
\label{eq80} \\ & & \qquad \times
{\rm tr}_\gamma \left[ iS^{f_1}(i\omega_n, \vec q) i \gamma_5
i S^{f_2}(i\omega_n-i\alpha_l, \vec q - \vec p) i \gamma_5
i S^{f_3}(i\omega_n-i\nu_m, \vec q - \vec k) \right] \nonumber
\quad,
\end{eqnarray}
where $(i\nu_m, \vec k)$ is the four momentum of the incoming scalar meson
and $(i\alpha_l, \vec p)$ the four momentum of one of the outgoing mesons.
It is once again understood that the complex meson frequencies are to
be analytically continued at the end of the calculation.
Taking the spinor trace in Eq.~(\ref{eq80}) leads to the form
\begin{eqnarray}
& & \Gamma(i\nu_m,\vec k; i\alpha_l, \vec p)
= \frac{4N_c}{\beta} \sum_n \int \frac{d^3q}{(2\pi)^3}
\\ & & \qquad \qquad \times
\frac{A-B}
{\left[(i\omega_n+\mu_1)^2-E_1^2\right]
\left[(i\omega_n-i\alpha_l+\mu_2)^2-E_2^2\right]
\left[(i\omega_n-i\nu_m+\mu_3)^2-E_3^2\right]} \nonumber
\end{eqnarray}
with
\begin{equation}
A = m_3 \vec q (\vec q - \vec p)
-m_2 \vec q (\vec q - \vec k)
+m_1 (\vec q - \vec p) (\vec q - \vec k) +m_1m_2m_3
\end{equation} \begin{eqnarray}
B &=& m_3(i\omega_n + \mu_1)(i\omega_n - i\alpha_l + \mu_2)
-m_2(i\omega_n + \mu_1)(i\omega_n - i\nu_m + \mu_3)
\nonumber \\ & & \quad +
m_1(i\omega_n - i\alpha_l + \mu_2)(i\omega_n - i\nu_m + \mu_3)
\end{eqnarray}
and for which the abbreviations $E_1 = \sqrt{\vec q ^2 + m_1^2}$, $E_2 =
\sqrt{(\vec q - \vec p)^2 + m_2^2}$ and $E_3 = \sqrt{(\vec q - \vec k)^2
+ m_3^2}$ have been introduced.

As was done in the case of the polarization, it is useful to make a
decomposition of this function in terms of elementary integrals
\begin{eqnarray}
\Gamma(i\nu_m,\vec k; i\alpha_l, \vec p) &=& -\frac{N_c}{8\pi^2} \Bigg[
(m_3-m_2) B_0(\vec k - \vec p, m_2, \mu_2, m_3, \mu_3, i\nu_m-i\alpha_l)
\nonumber \\ & & \quad +
(m_1-m_2) B_0(\vec p, m_1, \mu_1, m_2, \mu_2, i\alpha_l)
\nonumber \\ & & \quad +
(m_1+m_3) B_0(\vec k, m_1, \mu_1, m_3, \mu_3, i\nu_m)
\nonumber \\ & & \quad +
\bigg[ m_1^2(m_3-m_2) + m_2^2(m_1+m_3) + m_3^2(m_1-m_2) - 2 m_1m_2m_3
\nonumber \\ & & \qquad +
m_3\left(\vec p^2-(i\alpha_l-\mu_2+\mu_1)^2\right)
 - m_2\left(\vec k^2-(i\nu_m-\mu_3+\mu_1)^2\right)
\nonumber \\ & & \qquad +
m_1\left((\vec p - \vec k)^2-(i\alpha_l-i\nu_m-\mu_2+\mu_3)^2\right) \bigg]
\nonumber \\ & & \quad \times
C_0(\vec p, \vec k, m_1, \mu_1, m_2, \mu_2, i\alpha_l, m_3, \mu_3, i\nu_m)
\Bigg]
\label{triangle} \quad.
\end{eqnarray}
The function $B_0$ has already been given in Eq.~(\ref{b0def}). It is
necessary to introduce a third loop integral $C_0$ (cf.
Appendix~\ref{app-loop}), which is explicitly given as
\begin{eqnarray}
& & C_0(\vec p, \vec k, m_1, \mu_1,
m_2, \mu_2, i\alpha_l, m_3, \mu_3,  i\nu_m) =
\frac{16\pi^2}{\beta} \sum_n e^{i\omega_n\eta}
\int\limits_{|\vec p|<\Lambda}
\frac{d^3q}{(2\pi)^3}
\label{eq88} \\ & & \qquad \times
\frac{1}{((i\omega_n+\mu_1)^2-E_1^2)}
\frac{1}{((i\omega_n-i\alpha_l+\mu_2)^2-E_2^2)}
\frac{1}{((i\omega_n-i\nu_m+\mu_3)^2-E_3^2)} \nonumber
\quad.
\end{eqnarray}
As with the polarization function that was given in in
Eq.~(\ref{polar}), Eqs.~(\ref{eq80}--\ref{triangle}) are defined
intentionally without any flavor factors.

\subsubsection{$t$ and $u$ channels}
The $t$ and $u$ channel exchange diagrams shown in Fig.~\ref{gengraph}
have the general form
\begin{eqnarray}
-i{\cal M}_t &=& f_t \delta_{c_1c_2} \bar v(p_2) i\gamma_5 (ig_1)
\frac{i}{p\! \! \! /_1  - p\! \! \! /_3  - m^{(t)}} i\gamma_5 (ig_2) u(p_1)
\nonumber\\
&=& i \frac{g_1g_2 f_t}{t-m^{(t)2}} \delta_{c_1c_2} \bar v(p_2) \gamma_5
(p\! \! \! /_1 - p\! \! \! /_3 + m^{(t)}) \gamma_5 u(p_1) \label{eq73}
\end{eqnarray} \begin{equation}
-i{\cal M}_u = i \frac{g_1g_2 f_u}{u-m^{(u)2}} \delta_{c_1c_2}
\bar v(p_2) \gamma_5 (p\! \! \! /_1 - p\! \! \! /_4  + m^{(u)})
\gamma_5 u(p_1) \label{eq74}
\end{equation}
where, once again $g_1$, $g_2$ are the quark--meson couplings for the
outgoing mesons, and $f_t$, $f_u$ account for flavor factors from the
Gell--Mann matrices.  The momenta of the incoming quark and antiquark
are $p_1$ and $p_2$, those of the outgoing mesons $p_3$ and $p_4$.  The
mass of the exchanged fermion is denoted by $m^{(t)}$ and $m^{(u)}$
respectively. Note that these masses are not necessarily equal.

For any actual calculation of cross sections and transition rates, one
needs to sum the relevant amplitudes, take the absolute value squared
and to sum over final and to average over initial states. Our
results are listed in Appendix~\ref{app-matrix}.

\subsection{Calculation of Cross Sections}
Having provided the prerequisites for calculation, we illustrate the
main features via examples. (i) The process $u\bar d\to \pi^+\pi^0$
provides an example in which a $u$ channel diagram is required. Such a
diagram occurs not only when the outgoing mesons are identical.
(ii) The process $u\bar s \to \pi^+K^0$ is chosen because it has
unequal masses for the incoming quarks, the virtual quarks of the
vertex part, and the outgoing mesons. (iii) The process $u\bar u\to
\pi^+\pi^-$ demonstrates the usage of mixing propagators for the
virtual scalar mesons in the $s$ channel. Together, these three examples
indicate how the calculation is performed in general.

We assume that the centre of mass system of the incoming quarks is at
rest relative to the medium. All quantities, such as couplings $g_{\pi
q\bar q}$, are calculated within this framework, and this is used in
all processes calculated, e.~g. in triangle diagrams.  Since this
implies that $\vec k=\vec 0$ in Eqs.~(\ref{eq80})--(\ref{eq88}), the
vertex function $\Gamma$ then depends only on the absolute value of the
meson momentum. This leads to the fact that the total cross section
depends only on the invariant energy $s$ and the temperature
$T$\cite{su2had}.

In what follows, we drop the momentum arguments of the scalar
meson propagator ${\cal D}$ and the three meson vertex $\Gamma$ for
simplicity.

\subsubsection{Calculation of $\sigma_{u\bar d \to \pi^+\pi^0}$}
\begin{figure} % Figure 9
\caption[]{Feynman graphs for the process $u\bar d\to\pi^+\pi^0$.}
      \label{ud-pipi}
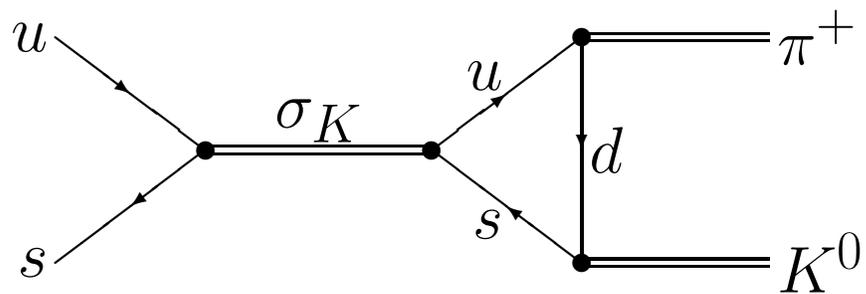
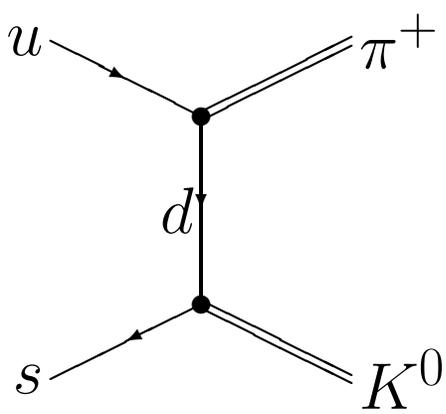
\end{figure}

Explicit graphs for this process are shown in Fig.~\ref{ud-pipi}.  We
have two graphs of the $s$ channel type, which we label $s$ and $s'$,
respectively.  The flavor factors for these graphs are found to be
\begin{mathletters}
\begin{equation}
f_s = -f_{s'} = -2\sqrt{2}
\end{equation} \begin{equation}
f_t = -f_u = -\sqrt{2}
\quad.
\end{equation}
\end{mathletters}
The relative sign of $f_s$, $f_{s'}$ and $f_t$, $f_u$ arises from the
flavor matrix $\lambda_3$ that occurs at the $\pi^0$ vertex, whose $uu$
and $dd$ components have opposite sign.

The virtual meson exchanged in this case is a charged $\sigma_\pi$, so
that one may identify
\begin{equation}
{\cal D}_s = {\cal D}_{s'} =
\frac{2K_1^-}{1-4K_1^-\Pi^S_{q\bar q}(\sqrt{s}, \vec 0)}
\quad.
\end{equation}
The contribution of the vertex graph in the centre of mass system turns
out in this case to be the same for both $s$ channel type graphs. From
our general formula Eq.~(\ref{triangle}), one has
\begin{eqnarray}
\Gamma_s = \Gamma_{s'} &=& - \frac{N_cm_q}{8\pi^2}\Big[
2B_0(\vec 0, m_q, \mu_q, m_q, \mu_q, \sqrt{s}) \label{eq94} \\
& & \qquad + (s-2m_\pi^2)
C_0(\vec p_3, \vec 0, m_q, \mu_q, m_q, \mu_q, \frac{1}{2}\sqrt{s},
m_q, \mu_q, \sqrt{s}) \Big] \nonumber
\quad,
\end{eqnarray}
where $|\vec p_3|=\frac{1}{2}\sqrt{s-4m_\pi^2}$ is the momentum of the
outgoing $\pi^+$.  The total contribution of the $s$ channel type graphs
is thus
\begin{eqnarray}
-i({\cal M}_s + {\cal M}_{s'}) &=& -i \bar v(p_2) u(p_1)\delta_{c_1c_2}
g_{\pi q\bar q}^2
\left(f_s {\cal D}_s\Gamma_s + f_{s'} {\cal D}_{s'}\Gamma_{s'}\right)
\nonumber \\
&=& 0 \quad,
\end{eqnarray}
which resembles the result for the two flavor model\cite{su2had},
where this process has no $s$ channel due to the lack of charged
scalar resonances.

The squared invariant amplitude thus arises from the $t$ and $u$ channels
alone. From Eqs.~(\ref{mattt}), (\ref{matuu}) and (\ref{mattu}),
this takes the form
\begin{eqnarray}
\frac{1}{4N_c^2} \sum_{s,c} \left| {\cal M}_t + {\cal M}_u \right|^2 &=&
\frac{|g_{\pi q\bar q}|^4}{N_c} \Bigg[
\frac{s(m_q^2-t)-(t-m_\pi^2-m_q^2)^2}{(t-m_q^2)^2}
\\ & & \qquad +
\frac{s(m_q^2-u)-(u-m_\pi^2-m_q^2)^2}{(u-m_q^2)^2}
\nonumber \\ & & \qquad -
2\frac{(t-m_\pi^2-m_q^2)(m_\pi^2+m_q^2-u)+sm_\pi^2}{(t-m_q^2)(u-m_q^2)}
\Bigg] \nonumber
\quad.
\end{eqnarray}
In the limit $m_\pi\to 0$, this reduces to an extremely simple
expression. This differs from that given in Ref.~\cite{su2had},
in which the $u$ channel has been omitted.

We obtain the differential cross section for this process to be
\begin{equation}
\frac{d\sigma_{u\bar d\to\pi^+\pi^0}}{dt} =
\frac{1}{16 \pi s (s-4m_q^2)} \frac{1}{4N_c^2}
\sum_{s,c} |{\cal M}_t+{\cal M}_u|^2
\quad.
\end{equation}

The total cross section is now constructed from the integration over
the exchanged momentum $t$ as\cite{su2had}
\begin{equation}
\sigma_{u\bar d\to\pi^+\pi^0}(s,T) = \int dt
\frac{d\sigma_{u\bar d\to\pi^+\pi^0}}{dt}
(1+f_B(\frac{1}{2}\sqrt{s}))^2
\quad.
\end{equation}
The Bose distribution function $f_B(x) = 1 / (\exp(\beta x) - 1)$
is introduced, since the presence of mesons in the heat bath leads
to an enhancement of the meson creation process in the medium.

\subsubsection{Calculation of $\sigma_{u\bar s \to \pi^+K^0}$}
\begin{figure} % Figure 10
\caption[]{Feynman graphs for the process $u\bar s\to\pi^+K^0$.}
      \label{us-pika}
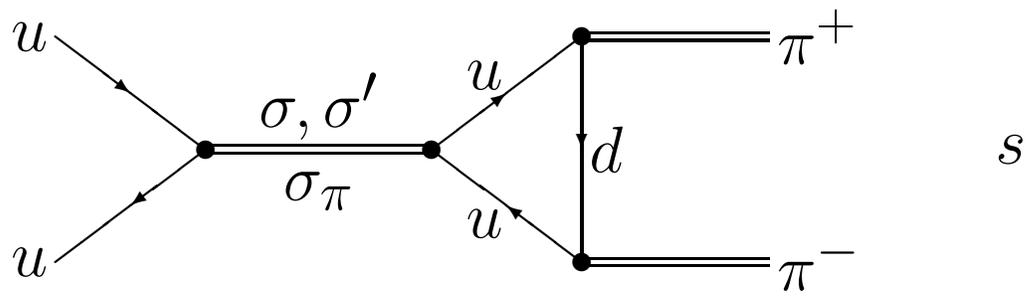
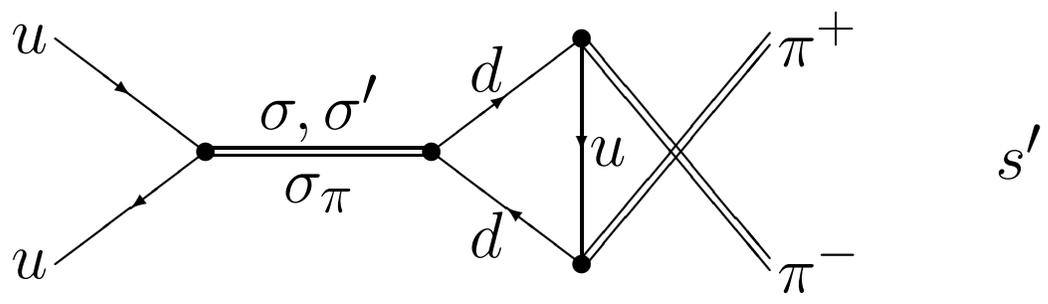
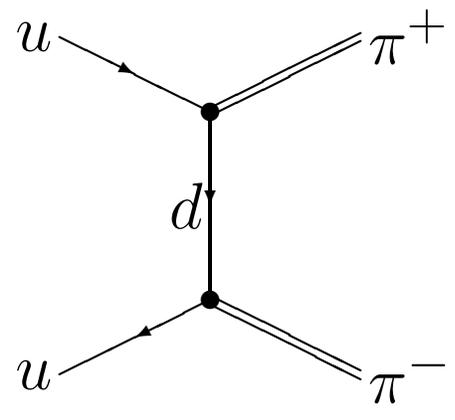
\end{figure}

The Feynman graphs for this process are shown in Fig.~\ref{us-pika}.
Here only one $s$ channel diagram is possible. From the graphs, we find
the flavor factors
\begin{mathletters}
\begin{equation}
f_s = 4
\end{equation} \begin{equation}
f_t = 2 \quad.
\end{equation}
\end{mathletters}
The virtual scalar meson exchanged in the s channel is a $\sigma_K$,
so that we have
\begin{equation}
{\cal D} = \frac{2K^-_4}{1-4K^-_4\Pi_{q\bar s}^S(\sqrt{s}, \vec 0)}
\quad.
\end{equation}
The contribution of the vertex part can be derived from our general
formula Eq.~(\ref{triangle}) to be
\begin{eqnarray}
\Gamma &=& - \frac{N_c}{8\pi^2}\Bigg[
(m_s-m_q)B_0(\vec p_4, m_q, \mu_q, m_s, \mu_s, E_4)
\nonumber \\ & & \qquad
+(m_s+m_q)B_0(\vec 0, m_q, \mu_q, m_s, \mu_s, \sqrt{s})
\nonumber \\ & & \qquad
-\bigg[ m_s m_\pi^2 + m_q\left(m_K^2-s+2E_3(\mu_s-\mu_q)\right) \bigg]
\nonumber \\ & & \qquad \qquad \times
C_0(\vec p_3, \vec 0, m_q, \mu_q, m_q, \mu_q, E_3, m_s, \mu_s, \sqrt{s})
\Bigg] \label{eq102}
\quad,
\end{eqnarray}
where $(E_3, \vec p_3)$ and $(E_4, \vec p_4)$ denote the four momenta of
the $\pi^+$ and the $K^0$, respectively.
Because the masses of the outgoing particles are different, the momentum
of the $\pi^+$ say, now has to be determined via
\begin{equation}
|\vec p_3| = \frac{1}{2\sqrt{s}}\sqrt{(s-(m_\pi+m_K)^2)
(s-(m_\pi-m_K)^2)} \quad. \label{pcms}
\end{equation}

{}From the formulae in Appendix~\ref{app-matrix}, one obtains the
squared transition amplitude as
\begin{eqnarray}
\frac{1}{4N_c^2}\sum_{s,c} |{\cal M}_s + {\cal M}_t|^2 &=&
\frac{8|g_{\pi q\bar q}g_{Kq\bar s}|^2}{N_c} \Bigg(
|{\cal D}\Gamma|^2\left(s-(m_q+m_s)^2\right)
\\ & & \qquad -
\frac{\Re({\cal D}\Gamma)}{t-m_q^2}\bigg( m_q(s-m_K^2+m_\pi^2)
\nonumber \\ & & \qquad \qquad +
(t-m_q^2-m_\pi^2)(m_q+m_s) \bigg)
\nonumber \\ & & \qquad +
\frac{1}{4(t-m_q^2)^2}\bigg((m_\pi^2+m_q^2-t)(t-m_K^2+m_s^2)
\nonumber \\ & & \qquad \qquad +
(m_q^2-t)(s-m_q^2-m_s^2)-2m_sm_qm_\pi^2\bigg)\Bigg)
\nonumber \quad,
\end{eqnarray}
with $\Re$ denoting the real part of the function following.

Due to the different masses of the $u$ and $s$ quarks, the differential
cross section is now given as
\begin{equation}
\frac{d\sigma_{u\bar s\to\pi^+K^0}}{dt} =
\frac{1}{64 \pi s \vec p_1^2} \frac{1}{4N_c^2}
\sum_{s,c} |{\cal M}_s+{\cal M}_t|^2
\quad.
\end{equation}
Again, the momentum $|\vec p_1|$ has to be calculated using
Eq.~(\ref{pcms}), where $m_\pi$, $m_K$ have to be replaced by $m_q$,
$m_s$, respectively.

The total cross section is defined to be
\begin{equation}
\sigma_{u\bar s\to\pi^+K^0}(s,T) = \int dt
\frac{d\sigma_{u\bar s\to\pi^+K^0}}{dt}
(1+f_B(E_3))(1+f_B(E_4))
\end{equation}
in this case.

\subsubsection{Calculation of $\sigma_{u\bar u \to \pi^+\pi^-}$}
The Feynman graphs for this process are shown in Fig.~\ref{uu-pipi}.
Once again, we have $s$ and $s'$ channel graphs. Since the incoming
quarks have the same flavor, we have to consider mixing for the
virtual mesons. The propagator for the $s$ graph reads
\begin{equation}
{\cal D}_s = M_{\sigma_\pi} + \frac{1}{3}
\left(2M_{00}+2\sqrt{2}M_{08}+M_{88}\right)
\quad,
\end{equation}
while the propagator for the $s'$ graph is
\begin{equation}
{\cal D}_{s'} = -M_{\sigma_\pi} + \frac{1}{3}
\left(2M_{00}+2\sqrt{2}M_{08}+M_{88}\right)
\quad.
\end{equation}
Here the flavor factors for the virtual meson vertices have already
been taken into account, so that we only have to consider the flavor
factors from the outgoing mesons, which contribute a factor $\sqrt{2}$
at every vertex:
\begin{equation}
f_s = f_{s'} = f_t = 2
\quad.
\end{equation}
The meson vertex part of the $s$ and $s'$ graph is the same as for
$u\bar d\to\pi^+\pi^0$, as was given in Eq.~(\ref{eq94}).  Since we
need the sum ${\cal M}_s + {\cal M}_{s'}$, we find that the
contribution of the $\sigma_\pi$ in this combination cancels and we
obtain
\begin{equation}
-i\left({\cal M}_s + {\cal M}_{s'}\right) =
-i\bar v(p_2)u(p_1)\delta_{c_1c_2}
g_{\pi q\bar q}^2\frac{4}{3}
\left(2M_{00}+2\sqrt{2}M_{08}+M_{88}\right)
\Gamma
\quad.
\end{equation}
The rest of the calculation proceeds as in the previous two cases
discussed.

\section{Numerical Results}
\label{resultsec}
In this section, we present our numerical results.  Since we are
interested in the cross sections as an input to transport equations, we
choose to give the transition rates $w(s,T)$ instead of the integrated
cross sections $\sigma(s,T)$. These two quantities are related
by\cite{deGroot}
\begin{equation}
w(s, T) = |\vec v_{\rm rel}| \sigma(s, T)
\quad, \label{rescale}
\end{equation}
where the relative velocity of the incoming particles in the centre of
mass system is
\begin{equation}
|\vec v_{\rm rel}| = \frac{\sqrt{(p_1p_2)^2-(m_1m_2)^2}}{E_1E_2}
\quad.
\end{equation}
For equal masses of the incoming quarks, $m_1=m_2=m$, this can be
expressed in terms of the centre of mass energy $s$ as
\begin{equation}
|v_{\rm rel}| = 2\sqrt{1-\frac{4m^2}{s}} \quad.
\end{equation}
The introduction of the transition rate as opposed to the cross
sections has a further advantage in that it suppresses the kinematic
singularity which appears at the threshold for the exothermic reactions
$q\bar q\to M_1M_2$ where $2m_q>m_{M_1}+m_{M_2}$.

To commence, we choose to discuss one simple process, and decompose the
calculation into its constituent $s$, $t$ and $u$ channel graphs, in
order to indicate the relative magnitude of these contributions.  We
pick the process $u\bar d\to\pi^+\eta$, that by Eq.~(\ref{sym2}) can
also be regarded as a component of the $\sigma_{u\bar u}$ calculation.
The appropriate Feynman diagrams can be obtained from Fig.~\ref{ud-pipi}
on replacing the $\pi^0$ by $\eta$ in this figure.  There are two $s$
channel, one $t$ channel and one $u$ channel exchanges possible. In this
case, we find $f_s=f_{s'}$ and $f_t=f_u$, so the $s$ channels do not
cancel.  We display these contributions explicitly in Fig.~\ref{single}.
Here we plot the transition rate that originates from $|{\cal M}_t|^2$
(which is equal to the contribution from $|{\cal M}_u|^2$), $|{\cal
M}_s|^2$, $|{\cal M}_{s'}|^2$, the sum $|{\cal M}_t+{\cal M}_u|^2$, and
the sum of all four graphs, $|{\cal M}_s+{\cal M}_{s'}+{\cal M}_t+{\cal
M}_u|^2$. From this figure, one can see that the $t$ and $u$ channels
give the dominant contributions.  Although the velocity factor in
Eq.~(\ref{rescale}) goes to zero at threshold, the $t$ channel
contribution stays finite.  However, it cancels against the $u$ channel
contribution, thus yielding a zero total rate here. This is a rather
general feature that we wish to point out.  Such a cancellation occurs
for all processes where the flavor factors of the $t$ and $u$ channel
graphs are equal, $f_t=f_u$.  We notice also that the $s$ channel
contributions are comparatively small, except in the region
$\sqrt{s}\approx 0.5\mbox{--}1\,{\rm GeV}$, where they exhibit a
resonance structure due to the $\sigma_\pi$ exchange. At threshold, their
contributions go to zero, since they are proportional to $s-4m_q^2$, as
can be seen explicitly from Eq.~(\ref{matss}). This demonstrates the
generic features that are empirically observed in all of our
calculations: in general the $t$ and $u$ channels dominate, while the
$s$ channel only plays a role over a small region if a resonance is
excited.

We turn now to a discussion of the four possible hadronization rates,
that stem from the four independent cross section classes $\sigma_{u\bar
d}$, $\sigma_{u\bar s}$, $\sigma_{u\bar u}$ and $\sigma_{s\bar s}$ that
were listed in Table~\ref{channeltab}.

\subsection{$u\bar d\to {\rm hadrons}$}
Figure~\ref{sigudchan} shows the hadronization rate of $u\bar d\to {\rm
hadrons}$ at $T=0$ as a function of $\sqrt{s}$. Here we have shown the
contributions from individual processes, in this case being $u\bar
d\to\pi^+\pi^0$, $K^+\overline{K^0}$ and $\pi^+\eta$.  The contribution
{}from the process $u \bar d \to \pi^+\eta'$ is negligible and has thus
not been shown. One recognizes from this figure that the transition rate
is dominated by the process $u\bar d\to \pi^+\pi^0$, as is expected
physically.  A further physical feature that is common to all the
calculations is also illustrated here, and we thus comment at this
point: that is, that an {\em exothermic\/} process displays a finite
rate at threshold. An exception to this rule follows for processes such
as $u\bar d\to\pi^+\eta$ (which are also exothermic), where equal flavor
factors occur in both the $t$ and $u$ channel exchanges and the amplitudes
cancel, as was discussed above.  By contrast, the process $u\bar d\to
K^+\overline{K^0}$ is {\em endothermic}, which means that the threshold
shifts from $\sqrt{s}=2m_q$ to $\sqrt{s}=2m_K$, where it yields a zero
rate. Finally we comment that the relative magnitude of the three rates
contributing to the total rate essentially reflects the available phase
space.

In Fig.~\ref{sigudt}, we show the hadronization rate at various values
of the temperature, $T=0$, $T=150\,{\rm MeV}$, $T=190\,{\rm MeV}$ and
$T=250\,{\rm MeV}$ as a function of $\sqrt{s}$. The threshold value of
$\sqrt{s}$ at which hadronization sets in is different for the four
cases due to the temperature dependence of the quark and meson masses.
At $T=0$ and $T=150\,{\rm MeV}$ the dominant process, $u\bar
d\to\pi^+\pi^0$, is exothermic, thus giving rise to a finite rate at
the threshold $\sqrt{s}=2m_q$.  At $T=190\,{\rm MeV}$ and $T=250\,{\rm
MeV}$ on the other hand, the pions are more massive than the quarks,
so that this process becomes endothermic and the threshold is given in
this case as $\sqrt{s}=2m_\pi$. Thus the finite values at thresholds
observed in the rates at lower temperatures change to become maxima
slightly above threshold, and a zero rate is obtained at these
thresholds.

In Fig.~\ref{sigdiff}, we show an example for the differential cross
section $d\sigma/d\cos\theta$ at zero temperature and at
$\sqrt{s}=1.5\,{\rm GeV}$.  Again the contribution of $u\bar
d\to\pi^+\eta'$ is negligible and is thus not shown. One notices a
forward--backward asymmetry in this plot, which is solely due to kaon
production.  Physically one can understand this as follows: after the
reaction $u\bar d\to K^+\overline{K^0}$ has occurred, the incoming $u$
quark has become a constituent of the $K^+$, so that the produced $K^+$
preferentially takes the direction of the incoming $u$.  In the other
reactions, $u\bar d\to\pi^+\pi^0$, $\pi^+\eta$ and $\pi^+\eta'$ however,
the $u$ can become a constituent of any one of the produced particles,
which makes these reactions forward--backward symmetric.

\subsection{$u\bar s\to {\rm hadrons}$}
Figure~\ref{siguschan} gives the transition rate for $u\bar s\to {\rm
hadrons}$ at $T=0$ as a function of $\sqrt{s}$, decomposed into its
constituent processes. The dominant contributions to the transition rate
come from the processes $u\bar s\to\pi^+K^0$ and $u\bar s\to\pi^0K^+$,
which give the same up to a factor of two. We again note that these are
exothermic processes, thus leading to a finite transition rate at the
threshold $\sqrt{s}=m_q+m_s$.  For $\sqrt{s}$ slightly above threshold,
a maximum appears in the transition rate of $u\bar s\to\pi^+K^0$,
$\pi^0K^+$, that corresponds to an excitation of the $\sigma_K$
resonance.  This maximum is not visible in the overall summed rate at
this temperature, since it is dominated rather by $u\bar s\to K^+\eta$
production.

Figure~\ref{sigust} shows the transition rates at the temperatures
$T=0$, $T=150\,{\rm MeV}$, $T=190\,{\rm MeV}$, $T=250\,{\rm MeV}$ as a
function of $\sqrt{s}$. The $\sigma_K$ excitation, that is not visible
at $T=0$, is now clearly seen at $T=150\,{\rm MeV}$ and $T=190\,{\rm MeV}$.
At $T=0$ and $T=150\,{\rm MeV}$, the threshold is determined as
$\sqrt{s}=m_q+m_s$, and the reaction is exothermic, while for
$T=190\,{\rm MeV}$ and $T=250\,{\rm MeV}$, the reactions are endothermic,
with threshold value $\sqrt{s}=m_\pi+m_K$.  Correspondingly, in the
exothermic regime, the transition rate is finite at threshold, while at
higher temperatures, where the outgoing mesons have become more massive
than the incoming quark and antiquark, the transition rate vanishes at
the threshold value.

\subsection{$u\bar u\to {\rm hadrons}$}
The contributions of the various processes for $u\bar u\to {\rm
hadrons}$ at $T=0$ are shown in Fig.~\ref{siguuchan} as a function of
$\sqrt{s}$.  Here we plot only the transition rates for the production
of $\pi^+\pi^-$, $\pi^0\pi^0$, $K^+K^-$ and $\pi^0\eta$. All other
channels are negligible.  One notices that the transition rate is
dominated by $\pi^+\pi^-$ production. For $\sqrt{s}> 1\,{\rm GeV}$, the
$K^+K^-$ production appears to be the next most important process,
followed by $\pi^0\pi^0$ and $\pi^0\eta$. We observe that the
contribution of $u\bar u\to\pi^0\pi^0$ is rather small in comparision
with $u\bar u\to\pi^+\pi^-$. We can trace this back to the allowed
Feynman diagrams and the associated flavor algebra. It turns out that
the $s$ channel exchanges for both processes give the same contribution
in total, although they have differing flavor factors associated with them.
On the other hand, the $t$ channel exchange for $\pi^0\pi^0$
contributes one half that obtained for $\pi^+\pi^-$ hadronization, due
to the associated flavor factors ($f_t=1$ for $u\bar u\to\pi^0\pi^0$,
$f_t=2$ for $u\bar u\to\pi^+\pi^-$). In addition, the $u\bar
u\to\pi^0\pi^0$ also has an $u$ channel exchange available. This tends
to cancel the $t$ channel exchange. Finally, the $t$ integration runs
only over half of the available phase space for the process $u\bar
u\to\pi^0\pi^0$\cite{su2had}. These features result in a considerable
reduction in the production of $\pi^0\pi^0$ over $\pi^+\pi^-$.  We also
point out that the production of $K^0\overline{K^0}$ is negligible in
contrast to the production of $K^+K^-$. This is due to the fact that
$u\bar u\to K^0\overline{K^0}$ proceeds only via a $s$ channel graph,
which gives small contributions compared to the $t$ and $u$ channels, as
has already been discussed.  This is a direct simulation of the Zweig
rule in the NJL model.

One final comment to the structures observed in Fig.~\ref{siguuchan} is
with regard to the cusp seen at threshold.  This is due to the
excitation of the $\sigma$ mode with $m_\sigma\approx 2m_q$.  At
$\sqrt{s}\approx 1.2\,{\rm GeV}$, one notices a very weak second
maximum: this comes from the $\sigma'$, appearing in the processes
$u\bar u\to K^+K^-$ and $u\bar u\to K^0\overline{K^0}$.

Figure~\ref{siguut} gives the rate for $u\bar u\to \mbox{\rm hadrons}$
at $T=0$, $T=150\,{\rm MeV}$, $T=190\,{\rm MeV}$ and $T=250\,{\rm MeV}$
as a function of $\sqrt{s}$. The cusp at the threshold value of
$\sqrt{s}$ at $T=0$ and $T=150\,{\rm MeV}$ develops to become a sharply
pronounced peak. This is a reflection of the exothermic nature of the
lower temperature processes, which have threshold $\sqrt{s}=2m_q$, as
opposed to the endothermic processes, with threshold $\sqrt{s}=2m_\pi$.
In this figure, one can see the $\sigma$ and $\sigma'$ maxima and the
shifting of the threshold due to the temperature dependence of the
masses.

\subsection{$s\bar s\to {\rm hadrons}$}
Finally we show the results for $s\bar s\to {\rm hadrons}$ for
different temperatures as a function of $\sqrt{s}$ in
Figs.~\ref{sigsschan} and \ref{sigsst}. Here the dominant contribution
comes from the processes $s\bar s\to K^0\overline{K^0}$ and $s\bar s\to
K^+K^-$, which have identical cross sections, and $s\bar s\to
\eta\eta$.  The pion production rate is very small, since these
processes do not have $t$ and $u$ channels available, again simulating
their suppression by the Zweig rule.  One notices a sharp resonance in
these curves near threshold. This is due to the $\sigma'$ intermediate
state. The threshold values of $\sqrt{s}$ are determined at $T=0$ and
$T=150\,{\rm MeV}$ by $\sqrt{s}=2m_s$, where the reaction is exothermic.
At $T=190\,{\rm MeV}$ and $T=250\,{\rm MeV}$, this is rather given by
$\sqrt{s}=2m_K$, since the reaction is endothermic.

\subsection{Averaged Transition Rates and Hadronization Times}
In order to make a connection with transport theory, we introduce
the energy averaged transition rate\cite{deGroot}
\begin{equation}
\bar w(T) = \frac{1}{\rho_1(T)\rho_2(T)}
\int \frac{d^3p_1}{(2\pi)^3}\frac{d^3p_2}{(2\pi)^3}
\left[2N_cf_F(E_1)\right] \left[2N_cf_F(E_2)\right] w(s,T)
\quad, \label{loss}
\end{equation}
which is the average transition rate for quarks and antiquarks coming
{}from a thermal medium, their distribution being given by the Fermi
function $f_F(x)=1/(\exp(\beta x)+1)$. The quark density is given as an
integral over this function
\begin{equation}
\rho_i(T) = \int \frac{d^3p}{(2\pi)^3}
2N_cf_F\left(\sqrt{\vec p^2+m_i^2}\right)
\label{dichte}
\end{equation}
for a given quark species.  In Eqs.~(\ref{loss}, \ref{dichte}), the
factor $2N_c$ accounts for the number of degrees of freedom.  To perform
the integration, we have made the approximation that $\sigma(s,T)$
depends only on $s$, even when we are not in the centre of mass system.
With this assumption, $\bar w(T)$ can be cast into the form
\begin{equation}
\bar w(T) = \int_{(m_1+m_2)^2}^\infty ds
\sqrt{(p_1p_2)^2-(m_1m_2)^2} \sigma(s,T) P(s,T) \quad,
\end{equation}
where the weight function $P(s,T)$ is given as
\begin{eqnarray}
P(s,T) &=& \frac{1}{\rho_1(T)\rho_2(T)}
\frac{1}{8\pi^4} \int_{m_1}^\infty dE_1 [2N_c f_F(E_1)] \\
& & \qquad \times
\int_{m_2}^\infty dE_2
[2N_c f_F(E_2)] \Theta(4|\vec p_1|^2|\vec p_2|^2
-(s-(m_1^2+m_2^2)-2E_1E_2)^2)
\nonumber \quad.
\end{eqnarray}

We show the results of the average rate $\bar w(T)$ for the four
hadronization classes $u\bar d$, $u\bar s$, $u\bar u$ and $s\bar s$ as a
function of temperature in Fig.~\ref{sigtot}. One observes that the
quantity $\bar w(T)$ stays fairly constant in the region from $T\approx
50\,{\rm MeV}$ up to the pion Mott temperature $T_{M\pi}$.  At
$T_{M\pi}$, it displays a minimum.  This behavior differs strikingly
{}from the curves shown in Ref.~\cite{su2had}:  whereas the cross
sections shown there decrease towards the critical temperature and then
{\em diverge\/} sharply, we obtain a {\em minimum\/} at the pion Mott
temperature with no divergence at all. There are two reasons for this
behaviour: (i) The reason for the occurrence of the divergence has been
explained in Ref.~\cite{su2had} as originating from the term
$d\sigma/dt\sim 1/(t-m_q^2)$ that arises from the $t$ channel
amplitude.  These authors have worked in the chiral limit, so that at
the critical temperature $m_q=0$, and the $t$ integration required for
the total cross section, yields a logarithmic divergence. Here, on the
other hand we consider the physical situation of {\em finite\/} current
quark masses, so that the quark propagators stay finite for all
temperatures, and no divergence may occur. (ii) The behaviour of the
cross sections is also determined by the quark--meson couplings, that
multiply the transition amplitude squared. In the chiral limit, as
calculated in Ref.~\cite{su2had}, these tend to a constant value at the
transition temperature, so that the overall behavior is given by the
mechanism described in (i). On the other hand, in our case, when
$m_0\ne 0$, the quark--meson couplings strictly approach zero at the
Mott temperatures, and are thus responsible for the observed behavior.

The little dip observed in the curves at $T=180\,{\rm MeV}$ corresponds
to the Mott temperature of the $\eta$, $T_{M\eta}$.  One also notes a
discontinuity in the production rate for the class $u\bar u\to \mbox{\rm
hadrons}$ at $T=165\,{\rm MeV}$. This occurs in the model since the
$\sigma$ in the $SU(3)\times SU(3)$ case is weakly {\em bound\/} at
$T=0$, and becomes a resonance at the Mott temperature
$T_{M\sigma}=165\,{\rm MeV}$. This leads to a transition rate $\sim
1/\sqrt{s-4m_q^2}$ and thus to a discontinuous averaged transition rate.
Due to kinematic reasons, this is only visible in the processes $u\bar
u\to\pi^+\pi^-$ and $u\bar u \to\pi^0\pi^0$ that contribute to $u\bar
u\to\mbox{\rm hadrons}$.  Note that this structure would not appear for
the $SU(2)\times SU(2)$ NJL model, where the $\sigma$ is a resonant
state for all temperatures.  This is thus a model dependent feature.

{}From $\bar w(T)$ we compute the hadronization times
\begin{equation}
\tau_u^{-1}(T) = \bar w_{u\bar u}(T)\rho_{\bar u} +
                      \bar w_{u\bar d}(T)\rho_{\bar d} +
                      \bar w_{u\bar s}(T)\rho_{\bar s}
\end{equation} \begin{equation}
\tau_s^{-1}(T) = \bar w_{s\bar u}(T)\rho_{\bar u} +
                      \bar w_{s\bar d}(T)\rho_{\bar d} +
                      \bar w_{s\bar s}(T)\rho_{\bar s}
\quad.
\end{equation}
The quantity $\tau_u(T)$ represents the mean lifetime of a $u$ quark in
a plasma before it hadronizes into a meson. Analogously $\tau_s(T)$ is
the mean lifetime of a $s$ quark. The numerical results are shown in
Fig.~\ref{tauhad}. In the temperature region of interest, $150\,{\rm
MeV}\le T\le 250\,{\rm MeV}$, the hadronization times are
$2\mbox{--}3\,{\rm fm}/c$ for a $u$ (or $d$) quark, and
$3\mbox{--}4\,{\rm fm}/c$ for a $s$ quark. As expected, the hadronization
time becomes infinite at the Mott temperature, since all hadronization
cross sections go to zero here.

\subsection{Strangeness Production}
In this section, we discuss the production rates for the processes that
change the total number of $s$ plus $\bar s$ quarks. The enhancement of
this quantity observed in the $K/\pi$ ratio in nucleus--nucleus
collisions over $N+N$ collisions is one of the observables which
indicates new physics\cite{qm91,KMR,str87}. The processes of interest
are (i) $u\bar d\to K^+\overline{K^0}$, (ii) $u\bar u\to K^+K^-$,
$K^0\overline{K^0}$ and (iii) $s\bar s\to\pi^+\pi^-$, $\pi^0\pi^0$,
$\eta\eta$, $\eta\eta'$, $\eta'\eta'$. While the first two reactions
{\em increase\/} the total number of $s+\bar s$, the third one {\em
decreases\/} this number.  The production/reduction rates $\bar w(T)$
are shown in Fig.~\ref{strtot}, respectively as a function of
temperature. Both increasing and decreasing processes show qualitatively
the same behavior.  The strangeness increasing processes are endothermic
and thus occur preferentially at higher temperatures, whereas the
strangeness decreasing processes (at least the dominant ones) are
exothermic and thus can also happen at low temperatures.  Since the
processes in these three cases do not dominate their respective
hadronization class, the overall magnitude of the rates is rather small
compared to that shown in Fig.~\ref{sigtot}.

It is useful to define the ratio
\begin{equation}
\frac{(\Delta K)_{\rm gain}}{\pi} = \frac{\bar w_{u\bar d\to K^+K^-} +
                             \bar w_{u\bar u\to K^+K^-} +
                             \bar w_{u\bar u\to K^0K^0}}
                            {\bar w_{u\bar d\to\pi^+\pi^0} +
                            2\bar w_{u\bar d\to\pi^+\eta} +
                             \bar w_{u\bar u\to\pi^+\pi^-} +
                             \bar w_{u\bar u\to\pi^0\pi^0} +
                            2\bar w_{u\bar u\to\pi^0\eta} +
                            3\bar w_{u\bar u\to\eta\eta}}
\end{equation}
measuring the total hadronization from up and down quarks into kaons
relative to the pion production due to hadronization. Here we have
assumed that the $\eta$ decays into three pions and we have neglected
the contribution of the $\eta'$. The temperature dependence of this
quantity is depicted in Fig.~\ref{enhance}. The discontinuity at
$T = 165\,{\rm MeV}$ is due to pion production, as has been
discussed previously. Kinks occur at the respective Mott temperatures
$T_{M\sigma}$, $T_{M\eta}$, $T_{MK}$ and $T_{M\pi}$. Numerically at
$T=150\,{\rm MeV}$, the strangeness enhancement ratio rises to $(\Delta
K)_{\rm gain}/\pi=0.01$. This enhancement of about 1\% is to be
compared with the enhancement of 9\% for $S+S$ collisions at $200\,{\rm
GeV}/A$ over $N+N$ collisions, as found by the NA35 collaboration
\cite{na35}. The loss term, defined as
\begin{eqnarray}
& &\frac{(\Delta K)_{\rm loss}}{\pi}= \\
& &\frac{\rho_s\rho_{\bar s} (\bar w_{s\bar s\to\pi^+\pi^-} +
                              \bar w_{s\bar s\to\pi^0\pi^0} +
                              \bar w_{s\bar s\to\eta\eta'} +
                              \bar w_{s\bar s\to\eta\eta} +
                              \bar w_{s\bar s\to\eta'\eta'})}
{2\rho_u\rho_{\bar u}(\bar w_{u\bar d\to\pi^+\pi^0} +
                     2\bar w_{u\bar d\to\pi^+\eta} +
                      \bar w_{u\bar u\to\pi^+\pi^-} +
                      \bar w_{u\bar u\to\pi^0\pi^0} +
                     2\bar w_{u\bar u\to\pi^0\eta} +
                     3\bar w_{u\bar u\to\eta\eta})}
\nonumber
\end{eqnarray}
is also indicated in Fig.~\ref{enhance}. The maximal loss is an order of
magnitude less than the gain in strangeness due to hadronization.  Note
that the back reaction $s\bar s\to\mbox{\rm nonstrange hadrons}$ has
been calculated under the assumption of thermal equilibrium values for
$\rho_s$ and $\rho_{\bar s}$. Should thermal equilibrium for strange
quarks however not be reached, one would expect this loss term to be
even more strongly suppressed.

\section{Summary and Conclusions}
\label{sumsec}
In this paper, we have used the $SU(3)\times SU(3)$ NJL model to
calculate temperature dependent cross sections and transition rates in
a scheme ordered in the inverse number of colors, $1/N_c$. To this end,
we have included an exact calculation of the pseudoscalar and scalar
meson masses, treating these particles as resonances  at temperatures
beyond their respective Mott temperatures, when they may
dissociate into their constituent quark and antiquark. Our calculation
has been performed for finite values of the current quark masses, which
is regarded as the physical situation.

In describing the hadronization procedure, we have identified four
classes $u\bar d$, $u\bar s$, $u\bar u$ and $s\bar s$, determined by the
initial incoming quark and antiquark and final states $\pi\pi$, $\pi K$,
$K\overline{K}$, $K\eta$ and $\eta\eta$. The individual processes
contributing to each class have been listed and calculated, and their
contributions to the cross sections have been studied as a function of
temperature and centre of mass energy. We are able to account for all
the observed structures, that result as a function of (i) threshold
effects, (ii) an intermediate resonance being excited, or via the (iii)
summation of several processes. The relevance of each process is
discussed, and we observe, for example, a forward--backward asymmetry in
the class $u\bar d\to\mbox{\rm hadrons}$, that is due to the process
$u\bar d\to K^+\overline{K^0}$. Averaged and total production rates are
calculated for each class.

What can be learned from this calculation? From the measured spectra of
the hadrons produced in high energy heavy ion collisions, one derives
temperatures in the range of $150\mbox{--}200\,{\rm MeV}$ \cite{na35}.
This temperature range is also found for the location of the phase
transition in lattice gauge calculations. In our calculation, the pion
Mott temperature $T_{M\pi}=212\,{\rm MeV}$ plays the role of separating
the regions where pions are stable ($T<T_{M\pi}$) and where they exist
as resonances. The unphysical aspect of our model is the appearance of
constituent quarks for $T<T_{M\pi}$. Thus all hadronization cross
sections for $T\ll T_{M\pi}$ are only of academic interest. If we focus
on the temperature region between 150\,MeV and 250\,MeV, two (in our
opinion) physically relevant numbers can be extracted.

(i) {\em The hadronization time:} The quark--gluon plasma (if it is
reached at all) is a transient state because of the rapid expansion of
the system. Therefore it is of great interest to know the times for the
various stages of the system: thermalization of the quark--gluon plasma,
hadronization and final state interaction time. In our calculation we
find an average time for the hadronization of a quark of
\begin{equation}
\tau_{\rm had} = 2 \mbox{--} 4 \, {\rm fm} / c \quad.
\end{equation}
This value may be considered not unrealistic in view of expansion
scenarios.

(ii) {\em Strangeness enhancement:} In experiments, a sizeable value of
strangeness enhancement $\Delta K/\pi$ has been observed in heavy ion
collisions over $N+N$ ones. This value is $\Delta K/\pi=9\%$. Our
calculation yields an enhancement of about 1\% as coming from the
hadronization stage. This result indicates that the dominant
contribution to strangeness enhancement must occur {\em before \/} or
{\em after\/} hadronization.

On the theoretical side, the present paper shows that hadronization
cross sections (as other quantities too) are nonperturbative in nature and
have a rather strong temperature dependence. Thus realistic calculations
(cascade or other ones) have to take these aspects into consideration,
if they want to be realistic.

\section*{Acknowledgments}
We wish to thank P.~Zhuang for illuminating discussions. This work has
been supported in part by the Deutsche Forschungsgemeinschaft under
contract no. Hu~233/4--3, by the Federal Ministry for Education and
Research, under contract no. 06~HD~742 and by the EU within the program
INTAS (94--2915).

\begin{appendix}
\section{Loop Integrals}
\label{app-loop}
The decomposition of Feynman diagrams into elementary integrals is a
well known technique for zero temperature\cite{nulloop}. At finite
temperature, however, no general theory has to date been given.
Large $SU(3)$ calculations are difficult without implementing a
technique, that enables one to decompose all expressions into a few
integrals only, the calculation of each of which can be done once.  The
evaluation of these integrals for the general case is a tedious task,
which will form the subject of a separate publication\cite{oneloop}. To
be concise, we simply illustrate the calculation of the elementary
integrals for some special cases here, confining ourselves to the
situation of zero chemical potentials and equal masses for all
fermions. This is already a sufficient basis for most of the
calculations within the two flavor sector.  A full calculation,
including differing chemical potentials, quark masses, and arbitrary
kinematics (applicable in $C_0$) can be found in a forthcoming
publication\cite{oneloop}.  Since the NJL model is nonrenormalizable,
we choose a three momentum cutoff $\Lambda$ as defining our
regularization scheme.

\subsection{Computation of $A$}
The first loop integral has been defined in Eq.~(\ref{Adef}) as
\begin{equation}
A(m, \mu) =
\frac{16\pi^2}{\beta} \sum_n e^{i\omega_n\eta}
\int\limits_{|\vec p| < \Lambda} \frac{d^3p}{(2\pi)^3}
\frac{1}{(i\omega_n+\mu)^2-E^2} \quad,
\end{equation}
where $E=\sqrt{p^2+m^2}$ and the limit $\eta\to 0$ has to be taken
after the Matsubara summation.  In the case $\mu =0$ this is easily
evaluated to be
\begin{equation}
A(m,0) = -4 \int_0^\Lambda dp \frac{p^2}{E}
\tanh\left(\frac{\beta E}{2}\right)
\quad.
\end{equation}
In the limit $\beta\to\infty$, this can be evaluated analytically. For
finite temperatures, the integral has to be performed numerically.

\subsection{Computation of $B_0$}
The second loop integral $B_0$ is defined as the analytic continuation of
\begin{eqnarray}
& &B_0(\vec k, m_1, \mu_1, m_2, \mu_2, i\nu_m) = \\ & &
\frac{16\pi^2}{\beta} \sum_n e^{i\omega_n\eta}
\int\limits_{|\vec p| < \Lambda} \frac{d^3p}{(2\pi)^3}
\frac{1}{((i\omega_n+\mu_1)^2-E_1^2)}
\frac{1}{((i\omega_n-i\nu_m+\mu_2)^2-E_2^2)}
\nonumber
\end{eqnarray}
($E_1=\sqrt{\vec p^2+m_1^2}$, $E_2=\sqrt{(\vec p - \vec k)^2+m_2^2}$) to
the real axis. The
case $\vec k=\vec 0$, which we require for the determination of the meson
masses, is singular and has to be treated separately. For this case, one
obtains
\begin{eqnarray}
B_0(\vec 0, m, 0, m, 0, k_0) &=& 8 {\cal P} \! \! \! \int dp
\frac{p^2\tanh\left(\frac{\beta E}{2}\right)}{4E^2-k_0^2}
\\ & & + i\pi\Theta(k_0-2m)\tanh\left(\frac{\beta k_0}{4}\right)
\sqrt{1-\left(\frac{2m}{k_0}\right)^2}
\quad. \nonumber
\end{eqnarray}
Here ${\cal P}$ denotes the Cauchy principal value of the integral.

For finite values of $|\vec k|$, $B_0$ takes the form
\begin{eqnarray}
B_0(\vec k, m, 0, m, 0, k_0) &=& -\frac{1}{k} \int_0^\Lambda
dp \frac{p\tanh\left(\frac{\beta E}{2}\right)}{E} \Bigg(
\log\left|\frac{(k^2+2p|\vec k|)^2-4k_0^2E^2}
{(k^2-2p|\vec k|)^2-4k_0^2E^2}\right| \\
& & + i\pi\left(\Theta(2p|\vec k|-|k^2+2k_0E|)
-\Theta(2p|\vec k|-|k^2-2k_0E|)\right) \Bigg)
\nonumber \quad.
\end{eqnarray}
The imaginary part of this integral can be easily evaluated
analytically. Note that this is not covariant due to our
noncovariant regularization scheme.

\subsection{Computation of $C_0$}
$C_0$ has been defined to be the analytic continuation of
\begin{eqnarray}
& & C_0(\vec p, \vec k, m_1, \mu_1,
m_2, \mu_2, i\nu_m, m_3, \mu_3,  i\alpha_l) =
\frac{16\pi^2}{\beta} \sum_n e^{i\omega_n\eta}
\int\limits_{|\vec p|<\Lambda}
\frac{d^3q}{(2\pi)^3} \\ & & \qquad \times
\frac{1}{((i\omega_n+\mu_1)^2-E_1^2)}
\frac{1}{((i\omega_n-i\alpha_l+\mu_2)^2-E_2^2)}
\frac{1}{((i\omega_n-i\nu_m+\mu_3)^2-E_3^2)} \nonumber
\quad.
\end{eqnarray}
In the example presented here, we not only set $\mu=0$ and consider
equal masses, but also set $\vec k = 0$, $|\vec q| =
\frac{1}{2}\sqrt{s-4m_\pi^2}$, $i\alpha_l=\frac{1}{2}\sqrt{s}$ and
$i\nu_m=\sqrt{s}$. The result can be directly applied to pion
production. One finds
\begin{eqnarray}
& &C_0(\vec 0, \vec q, m, 0,
m, 0, \sqrt{s}, m, 0, \frac{1}{2}\sqrt{s}) =
-\frac{1}{\sqrt{s(s-4m_\pi^2)}} {\cal P} \! \! \! \int dp
\frac{p\tanh\left(\frac{\beta E}{2}\right)}{E}
\label{c0pion}
\\ & & \qquad \times \Bigg[
\frac{1}{E}\Bigg(
\log \left|\frac{m_\pi^4-\zeta_-^2}{m_\pi^4-\zeta_+^2}\right|
-i\pi\left(\Theta(p|\vec q|-|\xi_+|)
+\Theta(p|\vec q|-|\xi_-|)\right)\Bigg)
\nonumber \\ & & \qquad \quad
+\frac{2}{\sqrt{s}-2E}\left(
\log\left|\frac{m_\pi^2-\zeta_-}{m_\pi^2-\zeta_+}\right|
-i\pi\Theta(p|\vec q|-|\xi_-|)\right)
\nonumber \\ & & \qquad \quad
+\frac{2}{\sqrt{s}+2E}\left(
\log\left|\frac{m_\pi^2+\zeta_+}{m_\pi^2+\zeta_-}\right|
+i\pi\Theta(p|\vec q|-|\xi_+|)\right)
\Bigg]
\nonumber \\ & & \qquad
+i\pi\frac{\tanh\left(\frac{\beta\sqrt{s}}{4}\right)}{\sqrt{s(s-4m_\pi^2)}}
\Bigg(
\log\left|\frac{2m_\pi^2-(s-|\vec q|\sqrt{s-4m^2})}
{2m_\pi^2-(s+|\vec q|\sqrt{s-4m^2})}\right|
\nonumber \\ & & \hspace{5cm}
-i\pi\Theta(|\vec q|\sqrt{s-4m^2}-|2m_\pi^2-s|)\Bigg)
\nonumber
\end{eqnarray}
($\zeta_\pm = E\sqrt{s}\pm p\sqrt{s-4m_\pi^2}$,
$\xi_\pm=E\sqrt{s}\pm m_\pi^2$).
In the limit $m_\pi\to 0$, the $\Theta$ functions vanish identically
and Eq.~(\ref{c0pion}) can be greatly simplified to yield the
result of Ref.~\cite{su2had} for the three meson vertex.

For a full calculation of $B_0$ for arbitrary values of the quark
masses and chemical potentials, as well as $C_0$ for the same and
arbitrary kinematics, the reader is referred to Ref.~\cite{oneloop}.

\section{Squared Matrix Elements}
\label{app-matrix}
In this appendix, we give the technical details required for
calculating the squared transition amplitude with unspecified general
masses. One is required to average over initial states and and sum over
final states. From the former, one obtains a factor $1/4N_c^2$. The
color trace gives a factor $N_c$. The flavor trace is accounted for
explicitly by the factors $f_s$, $f_t$ and $f_u$.  Our final
expressions for these functions, after taking the spinor trace, are
\begin{equation}
\frac{1}{4N_c^2}\sum_{s,c}|{\cal M}_s|^2 = \frac{f_s^2|g_1g_2|^2}{2N_c}
|{\cal D}\Gamma|^2 (s-(m_1+m_2)^2) \label{matss}
\end{equation} \begin{eqnarray}
\frac{1}{4N_c^2}\sum_{s,c}|{\cal M}_t|^2 &=& \frac{f_t^2|g_1g_2|^2}{2N_c}
\frac{1}{(t-m^{(t)2})^2} \bigg[ (m_3^2-m_1^2-t)(t-m_4^2+m_2^2)
\label{mattt} \\ & & \qquad +
(m^{(t)2}-t)(s-m_1^2-m_2^2) + 2m_1m^{(t)}(t-m_4^2+m_2^2)
\nonumber \\ & & \qquad -
2m_2m^{(t)}(m_3^2-m_1^2-t) -2m_1m_2(t+m^{(t)2}) \bigg]
\nonumber
\end{eqnarray} \begin{eqnarray}
\frac{1}{4N_c^2}\sum_{s,c}|{\cal M}_u|^2 &=& \frac{f_u^2|g_1g_2|^2}{2N_c}
\frac{1}{(u-m^{(u)2})^2} \bigg[ (m_4^2-m_1^2-u)(u-m_3^2+m_2^2)
\label{matuu} \\ & & \qquad +
(m^{(u)2}-u)(s-m_1^2-m_2^2) + 2m_1m^{(u)}(u-m_3^2+m_2^2)
\nonumber \\ & & \qquad -
2m_2m^{(u)}(m_4^2-m_1^2-u) -2m_1m_2(u+m^{(u)2}) \bigg]
\nonumber
\end{eqnarray}
for the squares of the individual channels. Here we denote the masses
of the incoming quarks by $m_1$, $m_2$, the masses of the outgoing
mesons by $m_3$, $m_4$ and the mass of of the virtual fermion in the
$t$ and $u$ channel by $m^{(t)}$ and $m^{(u)}$ respectively.

The results for the mixed terms are
\begin{eqnarray}
\frac{1}{4N_c^2}\sum_{s,c}{\cal M}_s{\cal M}_t^* &=& -
\frac{f_sf_t|g_1g_2|^2}{2N_c}\frac{{\cal D}\Gamma}{t-m^{(t)2}} \bigg[
m_1(m_2^2-m_4^2+t) \label{matst}
\\ & & \qquad -
m_2(m_3^2-m_1^2-t)+
m^{(t)}(s-(m_1+m_2)^2)\bigg] \nonumber
\end{eqnarray} \begin{eqnarray}
\frac{1}{4N_c^2}\sum_{s,c}{\cal M}_s{\cal M}_u^* &=& -
\frac{f_sf_u|g_1g_2|^2}{2N_c}\frac{{\cal D}\Gamma}{u-m^{(u)2}} \bigg[
m_2(m_1^2-m_4^2+u) \label{matsu}
\\ & & \qquad -
m_1(m_3^2-m_2^2-u)+
m^{(u)}(s-(m_1+m_2)^2)\bigg] \nonumber
\end{eqnarray} \begin{eqnarray}
\frac{1}{4N_c^2}\sum_{s,c}{\cal M}_t{\cal M}_u^* &=&
\frac{f_tf_u|g_1g_2|^2}{4N_c}\frac{1}{(t-m^{(t)2})(u-m^{(u)2})}
\label{mattu} \\ & & \qquad
\times \bigg[
(m_2^2+m_3^2-u)(m_4^2-m_1^2+2m_1m^{(u)}-u)
\nonumber \\ & & \qquad \qquad +
(s-(m_1-m_2)^2)(m_3^2+m_4^2-s)
\nonumber \\ & & \qquad \qquad +
(m_2^2+m_4^2-t)(m_3^2-m_1^2+2m_1m^{(t)}-t)
\nonumber \\ & & \qquad \qquad +
2m_2(m_1-m^{(u)})(m_1^2+m_3^2-t)
\nonumber \\ & & \qquad \qquad +
2m_2(m_1-m^{(t)})(m_1^2+m_4^2-u)
\nonumber \\ & & \qquad \qquad +
2(m_1-m^{(u)})(m_1-m^{(t)})(s-(m_1+m_2)^2) \bigg] \nonumber \quad.
\end{eqnarray}
{}From these formulae, it is straightforward to derive the results of
Ref.~\cite{su2had} for the case $m_1=m_2=m^{(t)}=m^{(u)}$ and
$m_3=m_4=0$.
\end{appendix}

\begin{figure} % Figure 1
\caption[]{The $q\bar q$ scattering amplitude in the random phase
      approximation.}
      \label{scatgraph}
\end{figure} \begin{figure} % Figure 2
\caption[]{Feynman diagram for the irreducible pseudoscalar
      polarization function.}
      \label{polgraph}
\end{figure} \begin{figure} % Figure 3
\caption[]{Temperature dependence of the constituent quark masses. The
      solid line refers to the light quarks up and down, the dashed line
      to the strange quark.}
      \label{gap}
\end{figure} \begin{figure} % Figure 4
\caption[]{Temperature dependence of the pseudoscalar meson masses, as
      well as that of $2m_q$ and $m_q+m_s$. Respective Mott
      temperatures are indicated by the solid points.}
      \label{massplot}
\end{figure} \begin{figure} % Figure 5
\caption[]{Temperature dependence of the pion (solid line) and kaon
      (dashed line) coupling strengths.}
      \label{cplplot}
\end{figure}      \begin{figure} % Figure 11
\caption[]{Feynman graphs for the process $u\bar u\to\pi^+\pi^-$.}
      \label{uu-pipi}
\end{figure} \begin{figure} % Figure 12
\caption[]{Decomposition of the rate for $u\bar d\to\pi^+\eta$
      at $T=0$ into its individual contributions.  At the threshold, the
      $t$ channel contribution is finite, the vertical line indicating
      the position of the threshold.}
      \label{single}
\end{figure} \begin{figure} % Figure 13
\caption[]{Contributions of individual hadronization processes for $u\bar
      d\to {\rm hadrons}$ at $T=0$.  The contribution of the process
      $u\bar d\to\pi^+\eta'$ is negligible and is thus not shown. The
      rate for $u\bar d\to\pi^+\pi^0$ is finite at threshold, which is
      indicated by the vertical line.} \label{sigudchan}
\end{figure} \begin{figure} % Figure 14
\caption[]{Transition rates for $u\bar d\to {\rm hadrons}$ at
      various temperatures. Solid line: $T=0$, dashed line:
      $T=150\,{\rm MeV}$, dotted line: $T=190\,{\rm MeV}$,
      dot--dashed line: $T=250\,{\rm MeV}$. At $T=0$ and $T=150\,{\rm
      MeV}$, the vertical line indicates a finite rate at threshold.}
      \label{sigudt}
\end{figure} \begin{figure} % Figure 15
\caption[]{Differential hadronization cross section for $u\bar d\to
      {\rm hadrons}$ at $T=0$, $\sqrt{s}=1.5\,{\rm GeV}$.}
      \label{sigdiff}
\end{figure} \begin{figure} % Figure 16
\caption[]{Contributions of individual hadronization processes for
      $u\bar s\to {\rm hadrons}$ at $T=0$.  The contribution of the
      process $u\bar s\to K^+\eta'$ is negligible and is thus not shown.
      The contributions of the processes $u\bar s\to\pi^+K^0$ and $u\bar
      s\to\pi^0K^+$ are finite at threshold.}
      \label{siguschan}
\end{figure} \begin{figure} % Figure 17
\caption[]{Transition rates for $u\bar s\to {\rm hadrons}$ at
      various temperatures. Solid line: $T=0$, dashed line:
      $T=150\,{\rm MeV}$, dotted line: $T=190\,{\rm MeV}$,
      dot--dashed line: $T=250\,{\rm MeV}$. At $T=0$ and $T=150\,{\rm
      MeV}$, the vertical line indicates a finite rate at threshold.}
      \label{sigust}
\end{figure} \begin{figure} % Figure 18
\caption[]{Contributions of individual hadronization processes for
      $u\bar u\to {\rm hadrons}$ at $T=0$. Only the four dominant
      processes are shown. The contribution of the process
      $u\bar u\to\pi^+\pi^-$ is finite at threshold.}
      \label{siguuchan}
\end{figure} \begin{figure} % Figure 19
\caption[]{Transition rates for $u\bar u\to {\rm hadrons}$ at
      various temperatures. Solid line: $T=0$, dashed line:
      $T=150\,{\rm MeV}$, dotted line: $T=190\,{\rm MeV}$,
      dot--dashed line: $T=250\,{\rm MeV}$. At $T=0$ and $T=150\,{\rm
      MeV}$, the vertical line indicates a finite rate at threshold.}
      \label{siguut}
\end{figure} \begin{figure} % Figure 20
\caption[]{Contributions of the dominant individual hadronization
      processes for $s\bar s\to {\rm hadrons}$ at $T=0$.  The
      contributions of the processes $s\bar s\to K^+K^-$ and $s\bar s\to
      K^+K^-$ are finite at threshold, as indicated by the vertical
      line.}
      \label{sigsschan}
\end{figure} \begin{figure} % Figure 21
\caption[]{Transition rates for $s\bar s\to {\rm hadrons}$ at
      various temperatures. Solid line: $T=0$, dashed line:
      $T=150\,{\rm MeV}$, dotted line: $T=190\,{\rm MeV}$,
      dot--dashed line: $T=250\,{\rm MeV}$. At $T=0$ and $T=150\,{\rm
      MeV}$, the vertical line indicates a finite rate at threshold.}
      \label{sigsst}
\end{figure} \begin{figure} % Figure 22
\caption[]{Temperature dependence of the averaged transition rates
      $\bar w(T)$.  Solid line: $u\bar d\to {\rm hadrons}$, dashed line:
      $u\bar s\to {\rm hadrons}$, dotted line: $u\bar u\to {\rm
      hadrons}$, dot--dashed line: $s\bar s\to {\rm hadrons}$.}
      \label{sigtot}
\end{figure} \begin{figure} % Figure 23
\caption[]{Temperature dependence of hadronization times for light
      quarks (solid line) and strange quarks (dashed line). Note
      the broken scale on the abscissa.}
      \label{tauhad}
\end{figure} \begin{figure} % Figure 24
\caption[]{Temperature dependence of the averaged strangeness
      production/destruction rates. Solid line: $u\bar d\to \mbox{\rm
      strange hadrons}$, dashed line: $u\bar u\to \mbox{\rm strange
      hadrons}$, dot--dashed line: $s\bar s \to \mbox{\rm nonstrange
      hadrons}$.}
      \label{strtot}
\end{figure} \begin{figure} % Figure 25
\caption[]{Temperature dependence of the strangeness enhancement
      factor $(\Delta K)_{\rm gain}/\pi$ (solid line) and the
      strangeness reduction factor $(\Delta K)_{\rm loss}/\pi$ (dashed
      line).}
      \label{enhance}
\end{figure}

\begin{table}
\caption{The four independent incoming $q\bar q$ states and their
         associated outgoing two meson states.}
\label{channeltab}
\begin{tabular}{|c||c|c|c|c|}
\hline
incoming & $u\bar d$ & $u\bar s$ & $u\bar u$ & $s\bar s$ \\ \hline
outgoing & $\pi^+ \pi^0$ & $\pi^+ K^0$ & $\pi^+ \pi^-$ &
                                              $\pi^+ \pi^-$ \\
         & $K^+ \overline{K^0}$ & $\pi^0 K^+$ & $\pi^0 \pi^0$ &
                                              $\pi^0 \pi^0$ \\
         & $\pi^+ \eta$ & $\eta K^+$ & $K^+ K^-$ & $K^+ K^-$ \\
         & $\pi^+ \eta'$ & $\eta' K^+$ & $K^0 \overline{K^0}$ &
			$K^0 \overline{K^0}$ \\
         & & & $\pi^0 \eta$ & $\eta \eta'$ \\
         & & & $\pi^0 \eta'$ & $\eta \eta$ \\
         & & & $\eta \eta'$ & $\eta' \eta'$ \\
         & & & $\eta \eta$ & \\
         & & & $\eta' \eta'$ & \\
\hline
\end{tabular}
\end{table}

\end{document}